\documentclass[aps,preprint,showpacs,floats,epsf,epsfig,nofootinbib,12pt]{revtex4}
\usepackage{graphicx}% Include figure files
\usepackage{epsfig}
\usepackage[dvips]{color}
\usepackage{subfigure}

\def\beq{\begin{eqnarray}}
\def\eeq{\end{eqnarray}}

\begin{document}

\title{Study on the possible molecular states composed of $\Lambda_c\bar D^*$, $\Sigma_c\bar D^*$, $\Xi_c\bar D^*$ and $\Xi_c'\bar D^*$ in the Bethe-Salpeter frame based on the pentaquark states $P_c(4440)$, $P_c(4457)$ and $P_{cs}(4459)$ }

\vspace{1cm}

\author{ Hong-Wei Ke$^{1}$   \footnote{khw020056@tju.edu.cn}, Fang Lu$^{1}$, Hai Pang$^{1}$\footnote{panghai@126.com}, Xiao-Hai
Liu$^1$\footnote{xiaohai.liu@tju.edu.cn}  and
        Xue-Qian Li$^2$\footnote{lixq@nankai.edu.cn}
   }

\affiliation{  $^{1}$ School of Science, Tianjin University,
Tianjin 300072, China
\\
  $^{2}$ School of Physics, Nankai University, Tianjin 300071, China }

\vspace{12cm}

\begin{abstract}
The measurements on a few pentaquarks states $P_c(4440)$,
$P_c(4457)$ and $P_{cs}(4459)$ excite our new interests about their structures. Since the masses of $P_c(4440)$ and
$P_c(4457)$ are close to the threshold of $\Sigma_c\bar D^*$, in the earlier works, they were regarded as molecular states of $\Sigma_c\bar D^*$
with quantum numbers $I(J^P)=\frac{1}{2}(\frac{1}{2}^-)$ and $\frac{1}{2}(\frac{3}{2}^-)$, respectively.
In a similar way $P_{cs}(4459)$ is naturally considered as a $\Xi_c\bar D^*$ bound state with $I=0$. Within the Bethe-Salpeter (B-S) framework we systematically study the possible bound states of $\Lambda_c\bar D^*$, $\Sigma_c\bar D^*$, $\Xi_c\bar D^*$ and $\Xi_c'\bar D^*$. Our results indicate that $\Sigma_c\bar D^*$ can form a bound state with $I(J^P)=\frac{1}{2}(\frac{1}{2}^-)$, which corresponds to $P_c(4440)$. However for the $I(J^P)=\frac{1}{2}(\frac{3}{2}^-)$ system the attraction between $\Sigma_c$ and $\bar D^*$ is too weak to constitute a molecule, so $P_{c}(4457)$ may not be a bound state of $\Sigma_c\bar D^*$ with $I(J^P)=\frac{1}{2}(\frac{3}{2}^-)$. As $\Xi_c\bar D^*$ and $\Xi_c'\bar D^*$ systems we take into account of the mixing between $\Xi_c$ and $\Xi'_c$ and the eigenstets should include two normal bound states $\Xi_c\bar D^*$ and $\Xi_c'\bar D^*$ with $I(J^P)=\frac{1}{2}(\frac{1}{2}^-)$ and a loosely bound state $\Xi_c\bar D^*$ with $I(J^P)=\frac{1}{2}(\frac{3}{2}^-)$. The conclusion that two $\Xi_c\bar D^*$ bound states exist, supports the suggestion that the observed peak of $P_{cs}(4459)$ may hide two states $P_{cs}(4455)$ and $P_{cs}(4468)$. Based on the computations we predict a bound state $\Xi_c'\bar D^*$  with $I(J^P)=\frac{1}{2}(\frac{1}{2}^-)$ but not that with $I(J^P)=\frac{1}{2}(\frac{3}{2}^-)$. Further more accurate experiments will test our approach and results.


\end{abstract}

\maketitle

\section{Introduction}
In recent years several pentaquark states have been successively measured in experiment, which broaden our scope of view on hadron physics. In 2019 the LHCb collaboration \cite{Aaij:2019vzc} observed a  pentaquark state $P_c(4312)$ in $J/\psi p$ invariant mass spectrum and found two  narrow overlapping  peaks $P_c(4440)$ and
$P_c(4457)$ which hid in the structure of the former observed $P_c(4450)$ \cite{LHCb:2015yax}.  In 2021 two similar states $P_{cs}(4338)$ \cite{LHCb:2022jad} and $P_{cs}(4459)$ \cite{LHCb:2020jpq} were announced in the $J/\psi \Lambda$ channel by the LHCb collaboration and their masses and widthes are collected in table \ref{Tab:p1}. According to their masses and the channels where they were found, they are regarded as pentaquark states rather than conventional excited baryonic states.
In fact in 2003 a baryon was measured at LEPS \cite{Nakano:2003qx}
which was conjectured as a pentaquark state, however later the
allegation was negated by further more accurate experiments. On the theoretical aspect Gell-Mann clearly indicated possible existence of pentaquark in his first paper on the quark model \cite{Gell-Mann:1964ewy}. Later some theoretical works have been addressed to this topic \cite{Riska:1992qd,Riska:2010ena,Zou:2010tc,Jaffe:2003sg,Karliner:2003dt,Shuryak:2003zi,Cheng:2004cc,Cheng:2004ew}.
For these new states people have all reasons to doubt about their inner structure because
more constituents involved would bring up more possible combinations,
unlike the simplest $q\bar q$ and $qqq$ for meson and  baryon, respectively.
In fact until now most researchers do not agree with each other on the inner structures of those exotic states recently found in experiments \cite{Belle:2005lik,Belle:2007hrb,Belle:2004lle,LHCb:2021uow,LHCb:2021auc}. Those discoveries on pentaquarks
have stimulated vigorous discussions on their properties \cite{Chen:2022asf}. Indeed the theoretical exploration
is crucial for getting a better understanding on their structures in the quark model and obtaining
valuable information about non-perturbative physics. Definitely,
 more accurate data achieved from experiments would help theorists to make progress on the route.

Now let us turn to some concrete subjects where charm physics is referred because relatively larger database is collected nowadays. There have been many papers which address to these hidden charm pentaquark states \cite{Chen:2019bip,Liu:2019tjn,Xiao:2019aya,Liu:2019zvb,He:2019ify,Xiao:2019mst,Zhang:2019xtu,Wang:2019hyc,Xu:2019zme,
Wang:2019got,Chen:2019asm,Lin:2019qiv,Cheng:2019obk,Fernandez-Ramirez:2019koa,Xu:2020gjl,Karliner:2022erb,Wang:2022mxy,Chen:2022wkh,Peng:2022iez,Azizi:2021pbh}.
For $P_c(4318)$ most authors suggest that it is a $\Sigma_c$ and $\bar D$ bound state. As for $P_c(4440)$ and $P_c(4457)$ they were regarded as the $\Sigma_c-\bar D^*$
molecular states with $J=\frac{1}{2}$ and $J=\frac{3}{2}$, respectively in the concerned papers \cite{Liu:2019tjn,He:2019ify,Xiao:2019mst,Chen:2019asm,Lin:2019qiv,Xu:2020gjl,Karliner:2022erb}. Certainly about  $P_c(4457)$ there are still some different opinions about its inner structure \cite{Peng:2022iez,Kuang:2020bnk}. The authors \cite{Wang:2022mxy,Karliner:2022erb} think $P_{cs}(4338)$ to be a $\Xi_c\bar D$ bound state, while they suggest that the observed $P_{cs}(4459)$ should correspond to two states $P_{cs}(4455)$ and $P_{cs}(4468)$ as $\Xi_c$$\bar D^*$ bound state with $I(J)=0(\frac{1}{2})$ and $0(\frac{3}{2})$, respectively. In that situation it requires further theoretical studies on the structures of $P_c(4440)$, $P_c(4457)$ and $P_{cs}(4459)$.

In Ref. \cite{Ke:2019bkf} we studied two possible molecular states $\Lambda_c\bar D$ and $\Sigma_c\bar D$ within the Bethe-Salpeter (B-S) framework. Our results indicate that the $\Sigma_c\bar D$ molecular state with $I=0$ can exist
which corresponds to the pentaquark state $P_c(4318)$. In this paper we still employ the same framework (B-S equation) to study the possible bound state of a charmed baryon ($\Lambda_c,\,\Sigma_c,\,\Xi_c$ or $\Xi'_c$) and $\bar D^*$. The
B-S equation is a relativistic equation  and established on the basis of quantum field
theory thus is adapted to deal with the bound
states \cite{Salpeter:1952ib}. Some authors have employed the B-S
equation to study the bound state of two
fermions \cite{Chang:2004im,Chang:2005sd}, the system of
one-fermion-one-boson \cite{Guo:1998ef,Weng:2010rb,Li:2019ekr,Wang:2019krq} and two bosons \cite{Guo:2007mm,Feng:2011zzb,Ke:2018jql,Ke:2012gm}.  In Ref. \cite{Xu:2020gjl}
the authors studied possible bound states of $\Sigma_c$
and $\bar D^*$ within their B-S framework. In this work we follow the approach adopted in \cite{Ke:2019bkf}  to
study the possible bound states of $\Sigma_c\bar D^*$,
$\Lambda_c\bar D^*$, $\Xi_c \bar D^*$ and $\Xi_c \bar D^*$. Though our  approach is a little similar to that in \cite{Xu:2020gjl} there is an evident difference  between them: we deduce the B-S equation and the corresponding kernel from the Feynman diagrams directly with the effective interactions; instead, in \cite{Xu:2020gjl} the authors directly wrote down  the potentials for exchanging different particles and then insert the kernel into these B-S equations.

The pentaquark states  $P_c(4440)$ and $P_c(4457)$ have been measured in the invariant
mass spectrum of $J/\psi p$  so their isospins are confirmed as $\frac{1}{2}$
because of isospin conservation. As for $P_{cs}(4459)$ the isospin is 0 since it is observed in $J/\psi \Lambda$ production. Thus we conjecture that the two
hadron constituents reside in an isospin eigenstate. For
the $\Lambda_c\bar D^*$ system  its isospin
must be $\frac{1}{2}$ but the $\Sigma_c\bar D^*$
system may reside in either isospin $\frac{1}{2}$ or
$\frac{3}{2}$. As for $\Xi_c\bar D^*$ and $\Xi'_c\bar D^*$ systems, the total isospin is 0 or $1$.  Certainly, for  a bound system composed of a  spin-parity
$\frac{1}{2}^+$ baryon and $1^-$ meson the total spin-parity is $\frac{1}{2}^-$ or $\frac{3}{2}^-$ if the two constituents are in the $S-$wave.

If two particles can mutually bind the interactions between them needs to be sufficiently strong. According to the quantum field theory two particles
interact via exchanging certain mediate particles. Since in our study two
constituents are color-singlet hadrons the
exchanged particles are some light hadrons such as $\pi$, $\eta$, $\sigma$, $\rho$ or (and)
$\omega$ etc..
The effective
interactions for the concerned heavy hadrons are deduced in Refs. \cite{Liu:2011xc,Colangelo:2005gb,Colangelo:2012xi,Ding:2008gr} and some of them we use
are collected in the appendix A.  Using the effective interactions we
obtain the corresponding B-S equation by the Feynman diagrams.

Inputting the corresponding parameters, one can solve the B-S equation numerically. For a
spin-isospin eigenstate, if the equation possesses a
solution with these reasonable parameters, then we would conclude that the corresponding bound state
could  exist in the nature, by contraries, no solution of the
B-S equation implies the supposed bound state cannot form.

This paper is organized as follows: after this introduction we
will derive the B-S equations related to  possible bound states
composed of a charmed baryon and $\bar D^*$.
Then in section III we will solve the B-S equation numerically and
present our  results by figures and tables. Section IV is devoted to a brief
summary.

\begin{table}
\caption{$P_c$ and $P_{cs}$ states observed by LHCb}\label{Tab:p1}
\begin{ruledtabular}
\begin{tabular}{cccc}
  state   & mass (MeV) &  width (MeV) \\\hline
  $P_c(4312)$  & $4311.9\pm 0.7^{+6.8}_{-0.6}$   &$9.8\pm 2.7^{+3.7}_{-4.5}$\\\hline
  $P_c(4440)$  & $4440.3\pm 1.3^{+4.1}_{-4.7}$   &$20.6\pm 4.9^{+8.7}_{-10.1}$\\\hline
  $P_c(4457)$  & $4457.3\pm 0.6^{+4.1}_{-1.7}$   &$6.4\pm 2.0^{+5.7}_{-1.9}$\\\hline
  $P_{cs}(4338)$  & $4338.2\pm 0.7\pm0.4$   &$7.0\pm 1.2\pm1.3$\\\hline
  $P_{cs}(4459)$  & $4458.8\pm 2.9^{+4.7}_{-1.1}$   &$17.3\pm 6.5^{+8.0}_{-5.7}$
\end{tabular}
\end{ruledtabular}
\end{table}

\section{The bound states of $\Lambda_c\bar D^*$,
$\Sigma_c\bar D^*$, $\Xi_c\bar D^*$ and $\Xi'_c\bar D^*$}

\begin{figure*}
        \centering
        \subfigure[~]{
          \includegraphics[width=6cm]{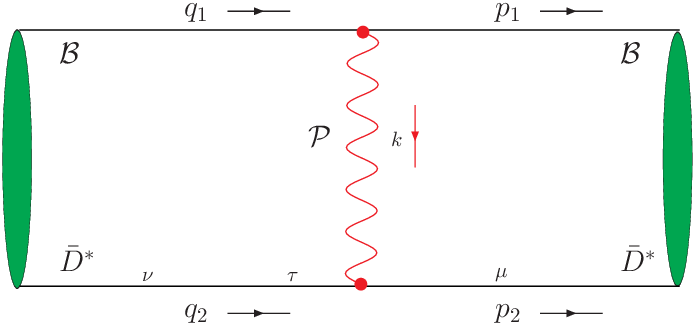}}\,\,\,\,\,\,\,\,\,\,\,\,\,\,\,\,\,\,\,\,\,\,
        \subfigure[~]{
          \includegraphics[width=6cm]{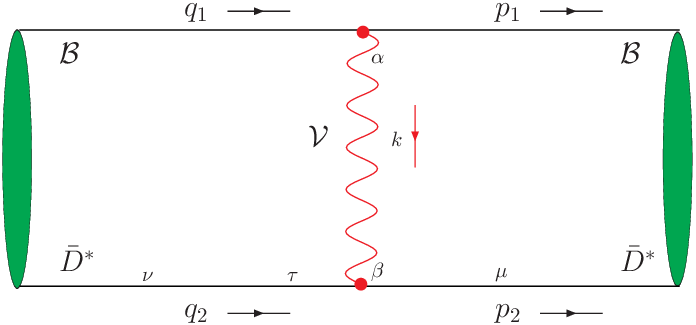}}
 \caption{the bound states of  $\mathcal{B}  \bar D^*$ formed by exchanging light pseudoscalar meson (a) or vector meson (b). }
        \label{penta1}
    \end{figure*}

Since the pentaquark states
$P_c(4440)$, $P_c(4457)$ and  $P_{cs}(4459)$ contain hidden
charms and their masses are close respectively to the
sums of the masses of single charmed baryons $\Lambda_c$,
$\Sigma_c$, $\Xi_c$ or $\Xi'_c$ and a charmed meson $\bar D^*$, it is obviously tempted to attribute those pentaquark states to
molecules of a charmed baryon and a charmed meson. Below
we will focus on the molecular
structures composed of a single charmed baryon and $\bar D^*$. Concretely, in this paper we
study $\Lambda_c\bar D^*$.
$\Sigma_c\bar D^*$, $\Xi_c\bar D^*$ and $\Xi'_c\bar D^*$ systems where the
spatial wave function between two constituents is in $S-$wave  i.e. its spin-parity ($J^P$) is $\frac{1}{2}^-$ or $\frac{3}{2}^-$.

\subsection{The isospin states of $\Lambda_c\bar D^*$ and $\Sigma_c\bar D^*$}

Since the isospins of $\Lambda_c$ and $\bar D^*$ are 0 and $\frac{1}{2}$, the isospin structure of the possible bound state
$\Lambda_c\bar D^{*}$ is
\begin{eqnarray}  \label{isospin1}
|\frac{1}{2},\frac{1}{2}\rangle=|\Lambda_c\bar D^{*0}\rangle.
\end{eqnarray}
 There also exists another state $|\frac{1}{2},-\frac{1}{2}\rangle=|\Lambda_c D^{*-}\rangle$ belonging to the doublet of $I=\frac{1}{2}$.

However the isospins of $\Sigma_c$  is 1, so the possible bound states of
$\Sigma_c\bar D^*$ could be in one of three isospin assignments i.e.
$|I,I_3\rangle$ are $|\frac{1}{2},\frac{1}{2}\rangle$
 $|\frac{3}{2},\frac{1}{2}\rangle$ and $|\frac{3}{2},\frac{3}{2}\rangle$. The explicit
isospin states are
\begin{eqnarray}  \label{isospin2}
|\frac{1}{2},\frac{1}{2}\rangle=\sqrt{\frac{2}{3}}|\Sigma_c^{++}
D^{*-}\rangle-\sqrt{\frac{1}{3}}|\Sigma_c^+\bar D^{*0}\rangle,
\end{eqnarray}
\begin{eqnarray}  \label{isospin3}
|\frac{3}{2},\frac{1}{2}\rangle=\sqrt{\frac{1}{3}}|\Sigma_c^{++}
D^{*-}\rangle+\sqrt{\frac{2}{3}}|\Sigma_c^+\bar D^{*0}\rangle,
\end{eqnarray}
and
\begin{eqnarray}  \label{isospin4}
|\frac{3}{2},\frac{3}{2}\rangle=|\Sigma_c^{++}\bar D^{*0}\rangle.
\end{eqnarray}
 Similarly  there are also three isospin adjoint states
$|\frac{1}{2},-\frac{1}{2}\rangle$,
$|\frac{3}{2},-\frac{1}{2}\rangle$  and
$|\frac{3}{2},-\frac{3}{2}\rangle$.

\subsection{The isospin states of $\Xi_c\bar D^*$ and $\Xi'_c\bar D^*$}

Since the isospins of $\Xi_c$ ($\Xi'_c$) and $\bar D^*$ are both 1/2, the total isospin  of the possible bound state of
$\Xi_c\bar D^{*}$ ($\Xi'_c\bar D^{*}$) is 0 or 1. For the $\Xi_c\bar D^{*}$ system  the explicit
isospin states are
\begin{eqnarray}  \label{isospin1}
|0,0\rangle=\frac{1}{\sqrt{2}}|\Xi^+_c D^{*-}\rangle -\frac{1}{\sqrt{2}}|\Xi^0_c D^{*0}\rangle \,\,,
\end{eqnarray}

\begin{eqnarray}  \label{isospin2}
|1,0\rangle=\frac{1}{\sqrt{2}}|\Xi^+_c D^{*-}\rangle +\frac{1}{\sqrt{2}}|\Xi^0_c D^{*0}\rangle ,
\end{eqnarray}
\begin{eqnarray}  \label{isospin3}
|1,1\rangle=|\Xi^+_c D^{*0}\rangle\rangle ,
\end{eqnarray}
and
\begin{eqnarray}  \label{isospin4}
|1,-1\rangle=|\Xi^0_c D^{*-}\rangle .
\end{eqnarray}

For the $\Xi'_c\bar D^{*}$ system one can replace $\Xi_c$ by  $\Xi'_c$ to obtain the isospin states.

\subsection{The Bethe-Salpeter (B-S) equation for $J^P=\frac{1}{2}^-$ and $\frac{3}{2}^-$ molecular states}
In the effective theory two hadrons interact via
exchanging light hadrons. For our concerned structures the Feynman diagrams at the leading order are
depicted in Fig. \ref{penta1} where $\mathcal{B}$, $\mathcal{P}$ and
$\mathcal{V}$  denote the charmed baryon, light pseudoscalar and vector mesons, respectively. For exchanging a light scalar meson (such as $\sigma$) the Feynman diagram is the same as Fig. \ref{penta1} (a), so we omit it.  The total and relative momenta of
the bound state are read as
\begin{eqnarray}\quad P = p_1 + p_2
=q_1 + q_2, \,\quad  p = \eta_2p_1 -
\eta_1p_2\,,\quad q = \eta_2q_1 - \eta_1q_2\,, \label{momentum-transform1}
\end{eqnarray}
where $P$ is the total momentum of the bound state, $q_1$ ($q_2$) and $p_1$ ($p_2$) are those
momenta of the constituents, $q$ and $p$ are the relative momenta at the two sides of the
effective vertex, $k$ is the
momentum of the exchanged meson, $\eta_i = m_i/(m_1+m_2)$ and
$m_i\, (i=1,2)$ is the mass of the $i$-th constituent meson.

The bound state composed of a baryon and a vector meson is written as
\begin{eqnarray} \label{4-dim-BS1}
\chi^\nu_P(x_1,x_2)
=\langle0|T\mathcal{B}(x_1)\mathcal{M}^\nu(x_2)|P\rangle.
\end{eqnarray}

The B-S wave function is a Fourier transformation of that into the momentum
space
\begin{eqnarray} \label{4-dim-BS3}
\chi^\nu_P(x_1,x_2) =e^{-iPX}\int\frac{d^4q}{(2\pi)^4}\chi^\nu_P(q)e^{-iqx}.
\end{eqnarray}

By the so-called ladder approximation the corresponding B-S
equation is deduced as
\begin{eqnarray} \label{4-dim-BS4}
\chi^\nu_{{P}}({ p})
=S_B(p_1)\int{d^4{q}\over(2\pi)^4}\,K_{\tau\mu}(P,p,q)\chi^\mu_{_{P}}(q)S^{\nu\tau}_M(p_2)\,,
\end{eqnarray}
where $S_B(p_1)$ is the propagator of the baryon ($\Lambda_c$,
$\Sigma_c$, $\Xi_c$ or $\Xi'_c$), $S^{\nu\tau}_M(p_2)$ is that of the meson ($\bar D^*$) and
$K_{\tau\mu}(P,p,q)$ is the kernel which is obtained by calculating the
Feynman diagram in Fig. 1. For the later convenience the
relative momentum $p$ is decomposed into the longitudinal $p_l$
($\equiv p\cdot v$) and transverse projection $p^\mu_t$ ($\equiv
p^\mu-p_lv^\mu$)=(0, $\mathbf{p}_T$) according to  the
momentum of the bound state $P$ ($v=\frac{P}{M}$). The two propagators are
\begin{eqnarray}\label{propagator1}
S_B(\eta_1
P+p)=\frac{i[(\eta_1M+p_l)v\!\!\!\slash+p_t\!\!\!\slash+m_1]}{[(\eta_1M+p_l)^2-\omega^2_1+i\epsilon]},
\end{eqnarray}
\begin{eqnarray}\label{propagator2}
S^{\mu\tau}_M(\eta_2
P-p)=\frac{i(-g^{\tau\mu}+p_2^\tau p_2^\mu/m_2^2)}{[(\eta_2M-p_l)^2-\omega^2_2+i\epsilon)]},
\end{eqnarray}
where $M$ is the total energy of the bound state, $\omega_i = \sqrt{ m_i^2-{
p_t}^2 }$ and $m_1$ ($m_2$) are the energy and the mass of the baryon (meson). Apparently the contribution of the tensor term in $S^{\nu\tau}_M$ is much smaller
than that of the first term for heavy meson, thus  it can be ignored in
numerical computations.

By the Feynman diagram the kernel $K_{\tau\nu}(P,p,q)$ is
written as

\begin{eqnarray}\label{kernel}
K_{\tau\nu}(P,p,q)=\sum_\mathcal{P} K^{\mathcal{P}}_{\tau\nu}(P,p,q)+\sum_\mathcal{S} K^{\mathcal{S}}_{\tau\nu}(P,p,q)+\sum_\mathcal{V} K^{\mathcal{V}}_{\tau\nu}(P,p,q),
\end{eqnarray}
with
\begin{eqnarray*}
&&K^{\mathcal{P}}_{\tau\nu}(P,p,q)=C_{{I\mathcal{P}}}g_{_{1\mathcal{P}}}g_{_{2\mathcal{P}}}\varepsilon_{\alpha\beta\mu\lambda}\gamma^\alpha\gamma^\beta k^\mu(p_1+q_2)^\lambda\varepsilon_{\theta\tau\nu\kappa}(p_2+q_2)^\theta k^\kappa\Delta(k,m_\mathcal{P})F^2(k),\\&&K^{\mathcal{S}}_{\tau\nu}(P,p,q)=C_{{I\mathcal{S}}}g_{_{1\mathcal{S}}}
g_{_{2\mathcal{S}}}g_{\tau\nu}\Delta(k,m_\mathcal{S})F^2(k),\\&&K^{\mathcal{V}}_{\tau\nu}(P,p,q)=C_{{I\mathcal{V}}}[g_{_{1\mathcal{V}}}(p_1+q_1)^\alpha-g'_{_{1\mathcal{V}}}
\gamma_\mu\gamma_\nu(k^\mu g^{\nu\alpha}-k^\nu g^{\mu\alpha})]
[-g_{_{2\mathcal{V}}}g_{\tau\nu}(p_2+q_2)^\beta\nonumber\\&&+g'_{_{2\mathcal{V}}}g_{\tau\theta}g_{\nu\delta}(k^{\theta}g^{\delta\beta}-k^{\delta}g^{\theta\beta})]\Delta_{\alpha\beta}(k,m_\mathcal{V})F^2(k),
\end{eqnarray*}
where $m_\mathcal{P}$ or $m_\mathcal{V}$ is the mass of the exchanged meson, $g_{_{1\mathcal{P}}}$, $g_{_{2\mathcal{P}}}$, $g_{_{1\mathcal{S}}}$, $g_{_{2\mathcal{S}}}$,
 $g_{_{1\mathcal{V}}}$, $g'_{_{1\mathcal{V}}}$, $g_{_{2\mathcal{V}}}$ and $g'_{_{2\mathcal{V}}}$
are the concerned coupling
constants, $C_{I\mathcal{M}}$ ($\mathcal{M}$ represent $\mathcal{P}$ where $\mathcal{S}$ or $\mathcal{V}$) are the isospin coefficients. We collect the isospin coefficients
in table \ref{Tab:p02} and \ref{Tab:p03} and one can refer to the appendix B of our early paper \cite{Ke:2019bkf} about how to obtain the isospin coefficients. The propagator of the exchanging mesnons are
$\Delta(k,m_\mathcal{M})=i/(k^2-m_\mathcal{M}^2)$ and
$\Delta_{\alpha\beta}(k,m_\mathcal{M})=i(-g_{\alpha\beta}+k_{\alpha}k_{\beta}/m_\mathcal{M}^2)/(k^2-m_\mathcal{M}^2)$.
In our study the exchanging particles are limited to only a few light mesons. Of course, exchanging two light mesons between
two hadrons may also induce a correction to the potential,
but it undergoes a loop  suppression, therefore, we do not
consider that contribution.
It is noted for $\Lambda_c \bar D^*$ system only  $\omega$ and $\sigma$ can be exchanged, however for $\Sigma_c \bar D^*$ system $\eta$, $\pi$,  $\sigma$, $\rho$ and $\omega$ can be
mediators. Simiary  for $\Xi_c \bar D^*$ system only $\rho$, $\omega$ and  $\sigma$ can be exchanged, but for $\Xi'_c \bar D^*$ system $\eta$, $\pi$,  $\sigma$, $\rho$ and $\omega$ can be
mediators. When we write down the B-S equation (similar for the transition matrix element) in terms of the Feynman diagrams we need to pay attention to the direction of the fermion line flow,  following the Feynman rules we put the B-S wave function of the final state in the left side of the equation and the  BS wave function of the initial state at the right side of the equation. The treatment is contrary to the common convention adopted for B-S equation.

\begin{table}
\caption{The isospin factors for $\Sigma_c\bar D^*$ systems
 }\label{Tab:p02}
\begin{ruledtabular}
\begin{tabular}{ccccccccccccc}
  $C_{\frac{1}{2}\pi}$   & $C_{\frac{1}{2}\eta}$& $C_{\frac{1}{2}\sigma}$  & $C_{\frac{1}{2}\rho}$   &  $C_{\frac{1}{2}\omega}$ & $C_{\frac{3}{2}\pi}$   & $C_{\frac{3}{2}\eta}$& $C_{\frac{3}{2}\sigma}$  & $C_{\frac{3}{2}\rho}$   &  $C_{\frac{3}{2}\omega}$ \\\hline
  -2  &1  &1 &-2  &1 &1&  1 &1  &1 &1
\end{tabular}
\end{ruledtabular}
\end{table}

\begin{table}
\caption{The isospin factors for  $\Xi'_c\bar D^*$ and (or) $\Xi_c\bar D^*$ systems
 }\label{Tab:p03}
\begin{ruledtabular}
\begin{tabular}{ccccccccccccc}
  $C_{0\pi}$   & $C_{0\eta}$& $C_{0\sigma}$  & $C_{0\rho}$   &  $C_{0\omega}$ & $C_{1\pi}$   & $C_{1\eta}$& $C_{1\sigma}$  & $C_{1\rho}$   &  $C_{1\omega}$ \\\hline
  -3  &1  &1 &-3  &1 &1&  1 &1  &1 &1
\end{tabular}
\end{ruledtabular}
\end{table}

Since the constituents of the molecule (meson and baryon) are not point particles,  a form
factor at each effective vertex should be introduced. The form factor is
suggested by many researchers as in the form:
\begin{eqnarray} \label{form-factor} F({\bf k},m_{\mathcal{ M}}^2 ) = {\Lambda^2 -
m_{ \mathcal{M}}^2 \over \Lambda^2 + {\bf k}^2}\,,\quad {\bf k} = {\bf
q}-{\bf p} \,,
\end{eqnarray}
where $\Lambda$ is a cutoff parameter which usually is taken as about 1 GeV.

The three-dimension B-S wave function is obtained after
integrating over $p_l$
\begin{eqnarray} \label{3-dim-BS1}
\chi^\mu_{{P}}({ p_t}) =\int\frac{dp_l}{2\pi}\chi^\mu_{{P}}({ p}).
\end{eqnarray}

For the $S-$wave system, the completely spatial wave function
can be found in Refs.\cite{Wang:2019krq,Weng:2010rb,Li:2019ekr}. However for the heavy hadrons case a simple version can be used \cite{Weng:2010rb}
\begin{eqnarray} \label{3-dim-BS2}
\chi^\mu_{{P}}({ p_t})
=[f_1(|\mathbf{p}_T|)+f_2(|\mathbf{p}_T|)p_t\!\!\!\slash]u^\mu(v,s),
\end{eqnarray}
where $f_1(|\mathbf{p}_T|)$ and $f_2(|\mathbf{p}_T|)$ are the
radial wave functions, $u^\mu(v,s)$, $v$ and $s$ are the spinors,
velocity and total spin of the bound state respectively. $u^\mu(v,s)=\frac{1}{\sqrt{2}}(\gamma^\mu+v^\mu)u(v,s)$ when the spin of the state is $\frac{1}{2}$ and $u^\mu(v,s)$
is Rarita-Schwinger vector spinor for a $J=\frac{3}{2}$ state.

In the following, let us substitute Eq. (\ref{kernel}) into Eq. (\ref{4-dim-BS4}) and employ the
so-called covariant instantaneous approximation where $q_l=p_l$
i.e. $p_l$ takes the place of $q_l$ in the kernel $K(P,p,q)$,
and then $K(P,p,q)$  no longer depends on $q_l$. Then we are performing a
series of manipulations: integrate over $q_l$ on the right side of
Eq. (\ref{4-dim-BS4}); multiply $\int\frac{dp_l}{(2\pi)}$ on the
both sides of Eq. (\ref{4-dim-BS4}), and integrate over $p_l$ on
the left side in the Eq. (\ref{4-dim-BS4}). Finally, substituting
Eq. (\ref{3-dim-BS2})  we obtain
\begin{eqnarray} \label{couple equation1}
&&[f_1(|\mathbf{p}_T|)+f_2(|\mathbf{p}_T|)p_t\!\!\!\slash]u^\mu(v,s)=\int\frac{dp_l}
{(2\pi)}\int\frac{d^3\mathbf{q}_T}{(2\pi)^3}\frac{
[(\eta_1M+p_l)v\!\!\!\slash+p_t\!\!\!\slash+m_1](g^{\mu\tau})}
{P_\mathcal{BM}}\nonumber\\&&\{\sum_\mathcal{P} K^{\mathcal{P}}_{\tau\nu}(P,p,q_t)+\sum_\mathcal{S} K^{\mathcal{S}}_{\tau\nu}(P,p,q_t)+\sum_\mathcal{V} K^{\mathcal{V}}_{\tau\nu}(P,p,q_t)\}
[f_1(|\mathbf{q}_T|)+f_2(|\mathbf{q}_T|)q_t\!\!\!\slash]u^\nu(v,s),
\end{eqnarray}
with
\begin{eqnarray*}\label{kernel2}
P_{\mathcal{BM}}=&&[(\eta_1M+p_l)^2-\omega^2_l+i\epsilon][(\eta_2M-p_l)^2-\omega^2_2+i\epsilon)],\\
K^{\mathcal{S}}_{\tau\nu}(P,p,q_t)=&&C_{{I\mathcal{S}}}g_{_{1\mathcal{S}}}g_{_{2\mathcal{S}}} \frac{ig_{\tau\nu}F^2(k)}{-(\mathbf{p}_T-\mathbf{q}_T)^2-m_\mathcal{S}^2},
\\K^{\mathcal{P}}_{\tau\nu}(P,p,q_t)=&&C_{{I\mathcal{P}}}g_{_{1\mathcal{P}}}g_{_{2\mathcal{P}}}\varepsilon_{\delta\theta\phi\omega}\gamma^\delta\gamma^\phi(p_{t}^\theta-q_{t}^\theta)[(2\eta_1P+2p_l)v^\omega+p^\omega_t+q^\omega_t] [(2\eta_2P-2p_l)v^\beta-p^\beta_t-q^\beta_t] \nonumber\\&&\frac{(p_t^\alpha-q_t^\alpha)\varepsilon_{\alpha\tau\nu\beta}iF^2(k)}{-(\mathbf{p}_T-\mathbf{q}_T)^2-m_\mathcal{P}^2},
\\K^{\mathcal{V}}_{\tau\nu}(P,p,q_t)=&&C_{{I\mathcal{V}}}\{[g_{_{1\mathcal{V}}}(2\eta_1P+2p_l)v^\alpha+p^\alpha_t+q^\alpha_t]+g'_{_{1\mathcal{V}}}\gamma_\omega\gamma_\phi[(p_{t}^\omega-q_{t}^\omega)g^{\phi\alpha}-(p_{t}^\phi-q_{t}^\phi)g^{\omega\alpha}]\}
\nonumber\\&&\{-g_{_{2\mathcal{V}}}g_{\tau\nu}[(2\eta_2P-2p_l)v^\beta-p^\beta_t-q^\beta_t] -g'_{_{2\mathcal{V}}}(g_{\tau\theta}g_{\nu\delta})[(p_{t}^{\theta}-q_{t}^{\theta})g^{\delta\beta}-(p_{t}^{\delta}-q_{t}^{\delta})g^{\theta\beta}]
\}\nonumber\\&&F^2(k)\frac{i[-g_{\alpha\beta}+(p_{t\alpha}-q_{t\alpha})(p_{t\beta}-q_{t\beta})/m_\mathcal{V}^2]}{{-(\mathbf{p}_T-\mathbf{q}_T)^2-m_\mathcal{V}^2}}.
\end{eqnarray*}

Now let us finally extract the expressions of $f_1(|\mathbf{p}_T|)$ and
$f_2(|\mathbf{p}_T|)$. Multiplying $\bar u^\mu(v,s)$ on both sides of
Eq.(\ref{couple equation1}), then by taking a trace, we get an expression which only
contains $f_1$ on the left side whereas multiplying $\bar u^\mu(v,s)p_t\!\!\!\slash$ to
the expression, $f_2$ is obtained on the left side, the
resultant formulaes are
\begin{eqnarray} \label{couple equation2}
&&f_1(|\mathbf{p}_T|)=\int\frac{dp_l}{(2\pi)}\int\frac{d^3\mathbf{q}_T}{(2\pi)^3}\frac{
i}
{P_{\mathcal{BM}}}[\sum_\mathcal{P}\frac{C_{{I\mathcal{P}}}g_{_{1\mathcal{P}}}g_{_{2\mathcal{P}}}F^2(k,m_\mathcal{P})K^\mathcal{P}_{1}}{-(\mathbf{p}_T-\mathbf{q}_T)^2-m_\mathcal{P}^2}+
\frac{C_{{I\mathcal{S}}}g_{_{1\mathcal{S}}}g_{_{2\mathcal{S}}}F^2(k,m_\mathcal{S})K^\mathcal{S}_{1}}{-(\mathbf{p}_T-\mathbf{q}_T)^2-m_\mathcal{S}^2}
\nonumber\\&&+\sum_\mathcal{V}C_{{I\mathcal{V}}}
\frac{F^2(k,m_\mathcal{V})(-g_{_{1\mathcal{V}}}g_{_{2\mathcal{V}}}K^{\mathcal{V},a}_{1}-g_{_{1\mathcal{V}}}g'_{_{2\mathcal{V}}}K^{\mathcal{V},b}_{1}-g'_{_{1\mathcal{V}}}g_{_{2\mathcal{V}}}K^{\mathcal{V},c}_{1}-g'_{_{1\mathcal{V}}}g'_{_{2\mathcal{V}}}K^{\mathcal{V},d}_{1})}{-(\mathbf{p}_T-\mathbf{q}_T)^2-m_\mathcal{V}^2}],
\end{eqnarray}
\begin{eqnarray} \label{couple equation3}
&&f_2(|\mathbf{p}_T|)=\int\frac{dp_l}{(2\pi)}\int\frac{d\mathbf{q}_T}{(2\pi)^3}\frac{
i}
{P_{\mathcal{BM}}}[\sum_\mathcal{P}\frac{C_{{I\mathcal{P}}}g_{_{1\mathcal{P}}}g_{_{2\mathcal{P}}}F^2(k,m_\mathcal{P})K^\mathcal{P}_{2}}{-(\mathbf{p}_T-\mathbf{q}_T)^2-m_\mathcal{P}^2}
+\frac{C_{{I\mathcal{S}}}g_{_{1\mathcal{S}}}g_{_{2\mathcal{S}}}F^2(k,m_\mathcal{S})K^\mathcal{S}_{2}}{-(\mathbf{p}_T-\mathbf{q}_T)^2-m_\mathcal{S}^2}
\nonumber\\&&+\sum_\mathcal{V}C_{{I\mathcal{V}}}\frac{F^2(k,m_\mathcal{V})(-g_{_{1\mathcal{V}}}g_{_{2\mathcal{V}}}K^{\mathcal{V},a}_{2}-g_{_{1\mathcal{V}}}g'_{_{2\mathcal{V}}}K^{\mathcal{V},b}_{2}-g'_{_{1\mathcal{V}}}g_{_{2\mathcal{V}}}K^{\mathcal{V},c}_{2}-g'_{_{1\mathcal{V}}}g'_{_{2\mathcal{V}}}K^{\mathcal{V},d}_{2})}{-(\mathbf{p}_T-\mathbf{q}_T)^2-m_\mathcal{V}^2}],
\end{eqnarray}
with
\begin{eqnarray*}
K^\mathcal{S}_{1}=&&C_{J\mathcal{S}}[ {f_1} ({\eta_1} M+p_l+m_1)-f_2{\mathbf{p}_T\cdot \mathbf{q}_T}],\\
K^\mathcal{S}_{2}=&& C_{J\mathcal{S}}[{f_1} +f_2({\eta_1} M+p_l-m_1){\mathbf{p}_T\cdot \mathbf{q}_T}/\mathbf{p}_T^2],\\
K^\mathcal{P}_{1}=&&\frac{16C_{J\mathcal{P}}}{3} {f_1} ({p_l}-{\eta_2} M) [({\eta_1} M+p_l)({\eta_1} M+p_l+m_1){(\mathbf{p}_T-\mathbf{q}_T)^2}+{\mathbf{p}_T\cdot \mathbf{q}_T}^2-{\mathbf{p}_T^2} {\mathbf{q}_T^2}]
\\&&+\frac{16C_{J\mathcal{P}}}{3}  {f_2}({p_l}-{\eta_2} M) [({\mathbf{p}_T\cdot \mathbf{q}_T}-{\mathbf{p}_T^2}) ({\mathbf{p}_T\cdot \mathbf{q}_T}-{\mathbf{q}_T^2}) ({\eta_1} M+{p_l})+{m_1} ({\mathbf{p}_T\cdot \mathbf{q}_T}^2-{\mathbf{p}_T^2}
   {\mathbf{q}_T^2})],\\
K^\mathcal{P}_{2}=&&\frac{16C_{J\mathcal{P}}}{3
   {\mathbf{p}_T^2}} {f_1} ({p_l}-{\eta_2} M) [({\mathbf{p}_T\cdot \mathbf{q}_T}-{\mathbf{p}_T^2})^2 ({\eta_1} M+{p_l})+{m_1} ({\mathbf{p}_T^2} {\mathbf{q}_T^2}-{\mathbf{p}_T\cdot \mathbf{q}_T}^2)]
\\&&-\frac{16C_{J\mathcal{P}}}{3 {\mathbf{p}_T^2}} {f_2} ({p_l}-{\eta_2} M)\{({\eta_1} M+p_l)({\eta_1} M+p_l-m_1)
   [{\mathbf{p}_T\cdot \mathbf{q}_T} ({\mathbf{p}_T^2}+{\mathbf{q}_T^2})-2 {\mathbf{p}_T^2} {\mathbf{q}_T^2}]\\&&+{\mathbf{p}_T^2}[{\mathbf{p}_T^2} {\mathbf{q}_T^2}-{\mathbf{p}_T\cdot \mathbf{q}_T}^2]\}
,\\
K^{\mathcal{V},a}_{1}=&&\frac{{f_1C_{Ja}}}{{m_\mathcal{V}}^2}({\eta_1} M+{m_1}+{p_l}) \{{m_\mathcal{V}}^2 [-4 {\eta_1} {\eta_2} M^2+4 M {p_l} ({\eta_1}-{\eta_2})+4 {p_l}^2-({\mathbf{p}_T}+{\mathbf{q}_T)^2}]-({\mathbf{p}_T^2}-{\mathbf{q}_T^2})^2\}\\&&+ {f_2C_{Ja}}{\mathbf{p}_T\cdot \mathbf{q}_T} [4 {\eta_1} {\eta_2} M^2+4 M {p_l} ({\eta_2}-{\eta_1})+\frac{({\mathbf{p}_T^2}-{\mathbf{q}_T^2})^2}{{m_\mathcal{V}}^2}-4 {p_l}^2+({\mathbf{p}_T}+{\mathbf{q}_T)^2}]
,\\
K^{\mathcal{V},b}_{1}=&&-\frac{4C_{Jb}}{3}{f_2} \left({\mathbf{p}_T\cdot \mathbf{q}_T}^2-{\mathbf{p}_T^2} {\mathbf{q}_T^2}\right)
,\\
K^{\mathcal{V},c}_{1}=&&4{f_1C_{Jc}} ({\mathbf{p}_T\cdot \mathbf{q}_T}-{\mathbf{p}_T^2}) ({p_l}-{\eta_2} M)\\&&+4{f_2C_J} \{[-{\eta_2} M  ({\eta_1} M+{m_1}+{p_l})+{\eta_1} M {p_l} +{m_1} {p_l}+{p_l}^2] ({\mathbf{p}_T\cdot \mathbf{q}_T}- {\mathbf{q}_T^2})-{\mathbf{p}_T\cdot \mathbf{q}_T}^2+{\mathbf{p}_T^2} {\mathbf{q}_T^2}\}
,\\
K^{\mathcal{V},d}_{1}=&&-\frac{8C_{Jd}}{3}{f_1} ({\mathbf{p}_T}-{\mathbf{q}_T)^2} ({\eta_1} M+{m_1}+{p_l})+\frac{8C_J}{3} {f_2} ({\mathbf{p}_T\cdot \mathbf{q}_T}-{\mathbf{p}_T^2}) ({\mathbf{p}_T\cdot \mathbf{q}_T}-{\mathbf{q}_T^2})
,\\
K^{\mathcal{V},a}_{2}=&&-\frac{f_1C_{Ja}}{{m_\mathcal{V}}^2}\{{m_\mathcal{V}}^2 [4 {\eta_1} {\eta_2} M^2+4 M {p_l} ({\eta_2}-{\eta_1})-4 {p_l}^2+({\mathbf{p}_T}+{\mathbf{q}_T)^2}]+({\mathbf{p}_T^2}-{\mathbf{q}_T^2})^2\}
\\&&
-\frac{f_2C_{Ja}}{{m_\mathcal{V}}^2 {\mathbf{p}_T^2}}{\mathbf{p}_T\cdot \mathbf{q}_T} (-{\eta_1} M+{m_1}-{p_l}) \{{m_\mathcal{V}}^2 [4 {\eta_1} {\eta_2} M^2+4 M {p_l} ({\eta_2}-{{\eta_1}})-4 {p_l}^2\\&&+({\mathbf{p}_T}+{\mathbf{q}_T)^2}]+({\mathbf{p}_T^2}-{\mathbf{q}_T^2})^2\}
,\\
K^{\mathcal{V},b}_{2}&&=\frac{f_2C_{Jb}}{3 {\mathbf{p}_T^2}}4 \left({\mathbf{p}_T\cdot \mathbf{q}_T}^2-{\mathbf{p}_T^2} {\mathbf{q}_T^2}\right) (-{\eta_1} M+{m_1}-{p_l})
,\\
K^{\mathcal{V},c}_{2}&&=-\frac{4f_1C_{Jc}}{{\mathbf{p}_T^2}} ({\mathbf{p}_T\cdot \mathbf{q}_T}-{\mathbf{p}_T^2}) ({p_l}-{\eta_2} M) (-{\eta_1} M+{m_1}-{p_l})\\&&-\frac{4 f_2C_{Jc}}{{\mathbf{p}_T^2}}\{M [{\eta_1} {\mathbf{p}_T\cdot \mathbf{q}_T}^2+{\eta_2} {\mathbf{p}_T\cdot \mathbf{q}_T} {\mathbf{p}_T^2}-{\mathbf{p}_T^2} {\mathbf{q}_T^2} ({\eta_1}+{\eta_2})]+{m_1}
   \left({\mathbf{p}_T^2} {\mathbf{q}_T^2}-{\mathbf{p}_T\cdot \mathbf{q}_T}^2\right)\\&&+{p_l} {\mathbf{p}_T\cdot \mathbf{q}_T} ({\mathbf{p}_T\cdot \mathbf{q}_T}-{\mathbf{p}_T^2})\}
,\\
K^{\mathcal{V},d}_{2}&&=-\frac{8f_1C_{Jd}}{3} ({\mathbf{p}_T}-{\mathbf{q}_T)^2}
-\frac{8f_2 C_{Jd}}{3 {\mathbf{p}_T^2}}({\mathbf{p}_T\cdot \mathbf{q}_T}-{\mathbf{p}_T^2}) ({\mathbf{p}_T\cdot \mathbf{q}_T}-{\mathbf{q}_T^2}) (-{\eta_1} M+{m_1}-{p_l})
\end{eqnarray*}
where the factors $C_{J\mathcal{P}}$, $C_{J\mathcal{S}}$, $C_{Ja}$, $C_{Jb}$, $C_{Jc}$ and $C_{Jd}$ are introduced to include the two cases of total spin: $J=$1/2 or 3/2, and their values are presented in table \ref{Tab:p4}.
\begin{table}
\caption{The factors $C_{J\mathcal{M}}$ where $\mathcal{M}$ represents $\mathcal{P}$, $\mathcal{S}$, $a$, $b$, $c$ and $d$.
 }\label{Tab:p4}
\begin{ruledtabular}
\begin{tabular}{ccccccccccccc}
  $C_{\frac{1}{2}\mathcal{P}}$   &$C_{\frac{3}{2}\mathcal{P}}$& $C_{\frac{1}{2}\mathcal{S}}$ & $C_{\frac{3}{2}\mathcal{S}}$  & $ C_{\frac{1}{2}a}$  &  $C_{\frac{3}{2}a} $& $C_{\frac{1}{2}b}$  & $ C_{\frac{3}{2}b} $   &   $C_{\frac{1}{2}c}$  &   $C_{\frac{3}{2}c}$  &   $C_{\frac{1}{2}d}$  &   $C_{\frac{3}{2}d}$  \\\hline
  1  &-1/2  &1 &1  &1 &1&  1 &-1/2  &1 &1&1&-1/2
\end{tabular}
\end{ruledtabular}
\end{table}

Now we  integrate over $p_l$ on the right side of Eqs.
(\ref{couple equation2}) and (\ref{couple equation3}) where four
poles exist at $-\eta_1M-\omega_1+i\epsilon$,
$-\eta_1M+\omega_1-i\epsilon$, $\eta_2M+\omega_2-i\epsilon$ and
$\eta_2M-\omega_2+i\epsilon$. By choosing an appropriate
contours  we
calculate the residuals at $p_l=-\eta_1M-\omega_1+i\epsilon$ and
$p_l=\eta_2M-\omega_2+i\epsilon$. The coupled equations after performing
contour integrations are obtained and collected in appendix (Eqs. (\ref{couple
equation12}) and (\ref{couple equation22}) ). One can carry out
the azimuthal integration and reduce Eqs. (\ref{couple equation12})
and (\ref{couple equation22}) to two one-dimensional integral equations
\begin{eqnarray}\label{one dimension equation}
&&f_1(|\mathbf{p}_T|)=\int{d|\mathbf{q}_T|}[A_{11}(|\mathbf{q}_T|,|\mathbf{p}_T|)
f_1(|\mathbf{q}_T|)+A_{12}(|\mathbf{q}_T|,|\mathbf{p}_T|)f_2(|\mathbf{q}_T|)],\\&&
f_2(|\mathbf{p}_T|)=\int{d|\mathbf{q}_T|}[A_{21}(|\mathbf{q}_T|,|\mathbf{p}_T|)
f_1(|\mathbf{q}_T|)+A_{22}(|\mathbf{q}_T|,|\mathbf{p}_T|)f_2(|\mathbf{q}_T|)],
\end{eqnarray}
where $A_{11}$, $A_{12}$, $A_{21}$ and $A_{22}$ are presented in
Appendix (see Eqs. (\ref{couple equation13}), (\ref{couple
equation14}), (\ref{couple equation23}) and (\ref{couple
equation24})).

\section{numerical results}

In order to solve the B-S equation numerically some parameters are
needed to be pre-set. The mass $m_{\Lambda_c}$, $m_{\Sigma_c}$, $m_{\Xi_c}$, $m_{\Xi'_c}$, $m_{D^*}$, $m_{\pi}$, $m_\sigma$,$m_\omega$,
$m_\rho$ are taken from the particle databook \cite{PDG18}. We follow
Ref. \cite{Liu:2011xc} to determine the parameters for the coupling constants of baryon-meson-baryon which are presented in table \ref{Tab:p5} and \ref{Tab:p6}. The coupling constants among mesons are collected in table \ref{Tab:p7} and Appendix A.

\begin{table}
\caption{The coupling constants of $\mathcal{BB}M$}\label{Tab:p5}
\begin{ruledtabular}
\begin{tabular}{cccccccccc}
  &$g_{1\pi}$  &  $g_{1\eta}$ & $g'_{1\sigma}$  &$ g_{1\rho}$& $g_{1\omega}$ &$ g'_{1\rho}$& $g'_{1\omega}$\\\hline
  $\Lambda_c \Lambda_c M$  &0  &0 &2$l_\mathcal{B}$ &0 &$\frac{\beta_{\mathcal{B}}g_{\mathcal{V}}}{2m_{\Lambda_c}}$& 0 &0 \\\hline
   $\Sigma_c \Sigma_c M$ &$\frac{g_1 }{4\sqrt{2}m_{\Sigma_c}f_\pi}$  &$\frac{g_1 }{4\sqrt{6}m_{\Sigma_c}f_\pi}$&$-l_\mathcal{S}$ &$ -\frac{\beta_{\mathcal{S}}g_{\mathcal{V}}}{4m_{\Sigma_c}}$&$ -\frac{\beta_{\mathcal{S}}g_{\mathcal{V}}}{4m_{\Sigma_c}}$ &$-\frac{\lambda_{\mathcal{S}}g_{\mathcal{V}}}{6}$&$-\frac{\lambda_{\mathcal{S}}g_{\mathcal{V}}}{6}$
   \\\hline
   $\Xi_c \Xi_c M$  &0  &0 &2$l_\mathcal{B}$ &$\frac{\beta_{\mathcal{B}}g_{\mathcal{V}}}{4m_{\Xi_c}}$ &$\frac{\beta_{\mathcal{B}}g_{\mathcal{V}}}{4m_{\Xi_c}}$&0 &0
   \\\hline
   $\Xi'_c \Xi'_c M$ &$\frac{g_1 }{8\sqrt{2}m_{\Xi'_c}f_\pi}$  &-$\frac{g_1 }{8\sqrt{6}m_{\Xi'_c}f_\pi}$&$-l_\mathcal{S}$ &$ -\frac{\beta_{\mathcal{S}}g_{\mathcal{V}}}{8m_{\Xi'_c}}$&$ -\frac{\beta_{\mathcal{S}}g_{\mathcal{V}}}{8m_{\Xi'_c}}$ &$-\frac{\lambda_{\mathcal{S}}g_{\mathcal{V}}}{12}$&$-\frac{\lambda_{\mathcal{S}}g_{\mathcal{V}}}{12}$
\end{tabular}
\end{ruledtabular}
\end{table}

\begin{table}
\caption{Some parameter in coupling constants}\label{Tab:p6}
\begin{ruledtabular}
\begin{tabular}{cccccccccc}
$g_1$ &$\beta_{\mathcal{B}}g_{\mathcal{V}}$ &$\beta_{\mathcal{S}}g_{\mathcal{V}}$&$\lambda_{\mathcal{S}}g_{\mathcal{V}}$(GeV$^{-1}$)&$l_\mathcal{B}$&$l_\mathcal{S}$  \\\hline1.0&-6.0&12&19.2&-3.65&7.3
\end{tabular}
\end{ruledtabular}
\end{table}

\begin{table}
\caption{The coupling constants of $\bar D^* \bar D^*M$}\label{Tab:p7}
\begin{ruledtabular}
\begin{tabular}{cccccccccc}
   & $g_{2\pi}$  & $g_{2\eta}$&$g'_{2\sigma}$&$g_{2\rho}$&  $g_{2\omega}$&$g'_{2\rho}$&  $g'_{2\omega}$\\\hline
  $D^* D^*M$  & $\frac{g_{\bar D^*\bar D*\mathcal{P}}}{2\sqrt{2}m_{D^*}}$  &$\frac{g_{\bar D^*\bar D*\mathcal{P}}}{2\sqrt{6}m_{D^*}}$  & $g_{\bar D^*\bar D*\mathcal{P}}$ &$\frac{g_{\bar D^*\bar D*\mathcal{V}}}{\sqrt{2}}$ &$\frac{g_{\bar D^*\bar D*\mathcal{V}}}{\sqrt{2}}$ &$\frac{g'_{\bar D^*\bar D*\mathcal{V}}}{\sqrt{2}}$ &$\frac{g'_{\bar D^*\bar D*\mathcal{V}}}{\sqrt{2}}$
\end{tabular}
\end{ruledtabular}
\end{table}

With these parameters, including the corresponding spin factors and the isospin factors, the coupled equations for the B-S wavefunction (Eqs. (\ref{one dimension
equation}) and (23)) are established. Since the coupled equations are
complicated integral equations, we shall solve them numerically.
The standard way is to discretize them, thus
one is able to convert them into algebraic  equations. Concretely,  within a reasonable
finite range (we set it from 0 to 2 GeV), we let $\bf |p_T|$ and $\bf |q_T|$ take $n$  discrete values
$Q_1$, $Q_2$,...$Q_n$ which distribute with equal gap from
$Q_1$=0.001 GeV to $Q_n$=2 GeV. The gap between two adjacent
values is $\Delta Q=1.999/(n-1)$ GeV (we set $n$=129 in our calculation). At this time the integral can be turned into summing the right sides of Eqs. \ref{one dimension
equation2}) and (25),

\begin{eqnarray}\label{one dimension equation2}
&&f_1(Q_i)=\sum_{j=1}^n \Delta Q [A_{11}(Q_j,Q_i)
f_1(Q_j)+A_{12}(Q_j,Q_i)f_2(Q_j)],\\&&
f_2(Q_i)=\sum_{j=1}^n \Delta Q [A_{21}(Q_j,Q_i)
f_1(Q_j)+A_{22}(Q_j,Q_i)f_2(Q_j)].
\end{eqnarray}
Since $Q_i$ can take the sequential values
$Q_1$, $Q_2$,...$Q_n$, the total number of the algebraic equations from Eqs. (\ref{one dimension
equation2}) and (25) is $2n$ and they can be written as a matrix equation

$$\left(\begin{array}{c}
      f_1(Q_1) \\... \\f_1(Q_{n})\\
        f_2(Q_1)\\... \\f_2(Q_{n})
      \end{array}\right)=A(\Delta E,\Lambda)\left(\begin{array}{c}
     f_1(Q_1) \\... \\f_1(Q_{n})\\
        f_2(Q_1)\\... \\f_2(Q_{n})
      \end{array}\right),$$
  where
$A(\Delta E,\Lambda)$ is a $2n\times 2n$
matrix whose elements are the coefficients of $f_1(Q_{j})$ and $f_2(Q_{j})$ given in Eqs. (\ref{one dimension
equation2}) and (25).

As a matter of fact, it is a homogeneous linear equation group. As long as
there exist non-trivial solutions, the necessary and sufficient
condition is that the coefficient determinant should be zero. In our case,
it is $|A(\Delta E,\Lambda)-I|=0$ where $I$ is the unit matrix and  $\Delta E=m_1+m_2-M$ is  the binding energy. $\Delta E$ and $\Lambda$ are two variables in the determinant. When  we set the value of $\Delta E$,
by requiring $|A(\Delta E,\Lambda)-I|=0$, we obtain a
corresponding $\Lambda$ if the equation group possesses solution. Generally $\Lambda$ should be close to 1 GeV. If the obtained $\Lambda$ is much beyond
the value or does not exist, we would conclude that the resonance cannot exist.
 With this strategy, we investigate the molecular
structure of $\Lambda_c$ and $\bar D^*$,
$\Sigma_c$ and $\bar D^*$, $\Xi_c$ and $\bar D^*$ as well as that of $\Xi'_c$ and $\bar D^*$.

\subsection{the numerical results on $\Lambda_c\bar D^*$ and $\Sigma_c\bar D^*$}

Now we begin  discussng the phenomenological consequences of the theoretical results. For $\Lambda_c$
and $\bar D^*$ system,  the isospin of $\Lambda_c$ is 0 and it
belongs to baryon anti-triplet of flavor ($\mathcal{B}_{\bar 3}$), so only $\omega$ and $\sigma$
can be exchanged between $\Lambda_c$ and $\bar D^*$. We cannot
find a solution for the B-S equation within a large range of
$\Lambda$ when we set some different binding energies $\Delta E$,
therefore we would conjecture  that the total interaction is repulsive between $\Lambda_c$ and $\bar D^*$.

With the same procedure, we study a molecule composed of
$\Sigma_c$ and $\bar D^*$. Since the isospin could be either $\frac{1}{2}$
or $\frac{3}{2}$ and the total spin could be either $\frac{1}{2}$
or $\frac{3}{2}$ , there are four possible states whose $I(J^P)$ are $\frac{1}{2}(\frac{1}{2}^-)$, $\frac{1}{2}(\frac{3}{2}^-)$, $\frac{3}{2}(\frac{1}{2}^-)$ and $\frac{3}{2}(\frac{3}{2}^-)$.  In this case $\pi$, $\eta$, $\sigma$, $\omega$ and
 $\rho$ exchanges between the two ingredients are allowed. For different binding energies the values of $\Lambda$ are presented in table \ref{Tab:p8} where the symbol ``-" means the value is larger than that for small $\Delta E$. We find for
 $I(J^P)=\frac{1}{2}(\frac{1}{2}^-)$ state when the binding energy is below 30 MeV the value of $\Lambda$
is close to 1, so a  $I(J^P)=\frac{1}{2}(\frac{1}{2}^-)$ state should exist.
However for  $I(J^P)=\frac{1}{2}(\frac{3}{2}^-)$  state even the binding energy is 1 MeV, the value of $\Lambda$
is much larger than 1, the fact means the abstractive interaction between $\Sigma_c$ and $\bar D^*$ with $I(J^P)=\frac{1}{2}(\frac{3}{2}^-)$ is too weak to form a bound state. For  $I(J^P)=\frac{3}{2}(\frac{1}{2}^-)$ and  $\frac{3}{2}(\frac{3}{2}^-)$   systems the B-S equations have no solution.

We study the interactions coming from different meson exchange  $I(J^P)=\frac{1}{2}(\frac{1}{2}^-)$ for $\Sigma_c\bar D^*$ state. When we ignore the interaction from $\pi$ in our calculation
the value of $\Lambda$ will increase to about 1.339 GeV with $\Delta E=1$ MeV. That means the interaction from $\pi$ exchange is attractive in this case. There is the same situation for $\rho$ exchange, i.e.
the interaction from $\rho$ exchange also is attractive. Since the isospin factor is opposite between $\rho$ exchange and $\omega$ exchange, but the interaction forms are same, the $\omega$ exchange
contributes a repulsive interaction. For $\Sigma_c \bar D^*$
molecule with $I(J^P)=\frac{1}{2}(\frac{1}{2}^-)$, the total interaction can be attractive due to a larger
contribution from the
$\rho$ and $\pi$ exchange.

{
 Apparently when $\Delta E$ is very
small the obtained $\Lambda$ is close to 1, so $\Sigma_c$ and
$\bar D^*$ should form a bound state with  $I(J^P)=\frac{1}{2}(\frac{1}{2}^-)$ . At present  the
pentaquark $P_c(4440)$ has been experimentally observed in
$\Lambda_b\to J/\psi pK$ channel, which peaks at the invariant
mass spectrum of $J/\psi p$ and has the invariant mass of about
4440.3 MeV. Apparently its isospin is $\frac{1}{2}$, and the
majority of authors
\cite{Liu:2019tjn,He:2019ify,Xiao:2019mst,Chen:2019asm,Lin:2019qiv,Xu:2020gjl,Karliner:2022erb}
regarded this pentaquark as a bound state of $\Sigma_c$ and $\bar
D^*$. We agree with it.} In experiment there is another state $P_c(4457)$ in company with $P_c(4440)$. Some authors suggest that $P_c(4457)$ should be
a bound state $\Sigma_c$ and $\bar
D^*$ with   $I(J^P)=\frac{1}{2}(\frac{3}{2}^-)$ . Our calculation indicates even for $\Delta E=1$ the value of $\Lambda$ is 1.927 GeV which deviates 1 GeV too much. The value of $\Lambda$ means the  attractive  interaction is very weak between two constituents so the system of $\Sigma_c\bar
D^*$ with  $I(J^P)=\frac{1}{2}(\frac{3}{2}^-)$ should not a resonance unless there is a special mechanism to enhance the  attractive  interaction .

By our observation given above, for the state with $I=\frac{3}{2}$
the isospin factor is 1 for  exchanging
either  $\omega$ or
$\rho$, therefore the total interaction is repulsive, it means that
$\Sigma_c$ and $\bar D^*$ cannot form a bound state
with $I=\frac{3}{2}$.

\subsection{the numerical results on $\Xi_c\bar D^*$ and $\Xi'_c\bar D^*$}

For the $\Xi_c \bar D^*$ system, if the other effects are not included  the corresponding $\Lambda$-values are presented in table \ref{Tab:p9}. One can find that even for a small binding energy the value
of $\Lambda$ is a bit too large. It seems $\Xi_c $ and $\bar D^*$ only may form two very loose bound states. In our calculation the two states $I(J^P)=0(\frac{1}{2}^-)$ and $0(\frac{3}{2}^-)$ are
also degenerated which is consistent with those in Ref. \cite{Wang:2022mxy}. Since only $\sigma$, $\rho$ and $\omega$ can be exchanged, the interactions are different only from the tensor item in the $\mathcal{L}_{{\bar \mathcal{D}^*\bar \mathcal{D}^*\mathcal{V}}}$ (see the values of $C_{\frac{1}{2}b}$ and $C_{\frac{3}{2}b}$ in table \ref{Tab:p4}), which indicates that
the tensor coupling has
little contribution to the interaction between $\Xi_c $ and $\bar D^*$.

For the $\Xi'_c \bar D^*$ system in terms of the original value of $\Lambda$ in table \ref{Tab:p10} one can find that the $I(J^P)=0(\frac{1}{2}^-)$ system can be a bound state but $I(J^P)=0(\frac{3}{2}^-)$ system
is not.

In Ref. \cite{Wang:2022mxy} with the coupled channel effect being taken into account  for their initial results, the values of $\Lambda$ apparently changed and the degeneration between the two states $I(J^P)=0(\frac{1}{2}^-)$ and $0(\frac{3}{2}^-)$ disappeared. In this article we employ another mechanism to achieve the same effect. We consider the effect of
the mixing between $\Xi_c$ and $\Xi'_c$ which has been discussed in some recent papers \cite{Ke:2022gxm,Geng:2022xfz,Liu:2023feb,Sun:2023noo}. Since the flavor symmetry of $SU(3)$ is broken the physical states $\Xi_c$ and $\Xi'_c$ should be
the mixing of $\Xi_c^{\bar 3}$ and $\Xi_c^{6}$ where the superscripts $\bar 3$ and $6$  correspond to $SU(3)_F$ anti-triple and sextet. Under the picture of mixing $\Xi_c^{\bar 3}$ and $\Xi_c^{6}$ should replace the $\Xi_c$ and $\Xi'_c$ in the representations $\mathcal{B}_{\bar 3}$ and $\mathcal{B}_6$ in Appendix A. By introducing a mixing angle $\theta$ the physical states $\Xi_c$ and $\Xi'_c$ can be expressed by $\Xi_c^{\bar 3}$ and $\Xi_c^{6}$ i.e. $\Xi_c={\rm cos}\theta\,\Xi_c^{\bar 3}+{\rm sin}\theta\,\Xi_c^{6}$ and $\Xi'_c=-{\rm sin}\theta\,\Xi_c^{\bar 3}+{\rm cos}\theta\,\Xi_c^{6}$. In Ref. \cite{Ke:2022gxm} we fix the value of
$\theta$ to be $16.27^\circ$ using the data of $\Gamma(\Xi_{cc}\to \Xi_c)/\Gamma(\Xi_{cc}\to \Xi'_c)$ . The results including the mixing effect are presented in table \ref{Tab:p11} and \ref{Tab:p12}. One can find

1. for the  $\Xi_c\bar D^*$ system
the values of $\Lambda$ drop but those for $\Xi'_c\bar D^*$ system raise;

2. for the  $\Xi_c\bar D^*$ system the degeneration of the two states $I(J^P)=0(\frac{1}{2}^-)$ and $0(\frac{3}{2}^-)$ disappear;

3. $\Xi_c\bar D^*$ should be form two states $I(J^P)=0(\frac{1}{2}^-)$ and $0(\frac{3}{2}^-)$ but $\Xi'_c\bar D^*$ only can form a $I(J^P)=0(\frac{1}{2}^-)$  bound state.

At present $P_{cs}(4459)$ was measured in experiment which is close to the threshold of $\Xi_c\bar D^*$ , so it seems $P_{cs}(4459)$ is one bound state of $\Xi_c\bar D^*$
with $I=0$. If the structure includes two states: $P_{cs}(4455)$ and $P_{cs}(4468)$ they are just corresponding to two states of  $\Xi_c\bar D^*$
with $I(J^P)=0(\frac{1}{2}^-)$ and $0(\frac{3}{2}^-)$,  it seems the suggestion in Refs. \cite{Wang:2022mxy,Karliner:2022erb} is reasonable. For $I=1$ systems there are no solutions can be obtained, so these states cannot exist.

\begin{table}
\caption{The cutoff parameter $\Lambda$ and the corresponding
binding energy $\Delta E$ for the bound state $\Sigma_c \bar D^*$
with $I=\frac{1}{2}$ }\label{Tab:p8}
\begin{ruledtabular}
\begin{tabular}{cccccccc}
  $\Delta E$ MeV  &1 & 5 &  10  &  20& 30 \\\hline
  $\Lambda (J=\frac{1}{2})$  &1.040 & 1.112    &1.166 &1.244   &1.306\\\hline
  $\Lambda (J=\frac{3}{2})$  &1.927 & 2.592    &3.617 &-   &-
\end{tabular}
\end{ruledtabular}
\end{table}

\begin{table}
\caption{The cutoff parameter $\Lambda$ and the corresponding
binding energy $\Delta E$ for the bound state $\Xi_c \bar D^*$
with $I=0$ }\label{Tab:p9}
\begin{ruledtabular}
\begin{tabular}{cccccccc}
  $\Delta E$ MeV  &1 & 5 &  10  &  20& 30 \\\hline
  $\Lambda (J=\frac{1}{2})$  &1.327 & 1.470    &1.568 &1.768   &1.924\\\hline
  $\Lambda (J=\frac{3}{2})$  &1.327 & 1.469    &1.585 &1.766   &1.923
\end{tabular}
\end{ruledtabular}
\end{table}
\begin{table}
\caption{The cutoff parameter $\Lambda$ and the corresponding
binding energy $\Delta E$ for the bound state $\Xi'_c \bar D^*$
with $I=0$ }\label{Tab:p10}
\begin{ruledtabular}
\begin{tabular}{cccccccc}
  $\Delta E$ MeV  &1 & 5 &  10  &  20& 30 \\\hline
  $\Lambda (J=\frac{1}{2})$  &1.070 & 1.141    &1.195 &1.271   &1.332\\\hline
  $\Lambda (J=\frac{3}{2})$  &1.800 & 2.341    &3.095 &-  &-
\end{tabular}
\end{ruledtabular}
\end{table}

\begin{table}
\caption{The cutoff parameter $\Lambda$ and the corresponding
binding energy $\Delta E$ for the bound state $\Xi_c \bar D^*$
with $I=0$ when the effect of the mixing between $\Xi_c$ and $\Xi'_c$ is included }\label{Tab:p11}
\begin{ruledtabular}
\begin{tabular}{cccccccc}
  $\Delta E$ MeV  &1 & 5 &  10  &  20& 30 \\\hline
  $\Lambda (J=\frac{1}{2})$  &1.144 & 1.235    &1.306 &1.411   &1.496\\\hline
  $\Lambda (J=\frac{3}{2})$  &1.252 & 1.381    &1.488 &1.657   &1.807
\end{tabular}
\end{ruledtabular}
\end{table}

\begin{table}
\caption{The cutoff parameter $\Lambda$ and the corresponding
binding energy $\Delta E$ for the bound state $\Xi'_c \bar D^*$
with $I=0$ when the effect of the mixing between $\Xi_c$ and $\Xi'_c$ is included}\label{Tab:p12}
\begin{ruledtabular}
\begin{tabular}{cccccccc}
  $\Delta E$ MeV  &1 & 5 &  10  &  20& 30 \\\hline
  $\Lambda (J=\frac{1}{2})$  &1.191 & 1.273    &1.336 &1.427   &1.545\\\hline
  $\Lambda (J=\frac{3}{2})$  &8.638 & -    &- &-   &-
\end{tabular}
\end{ruledtabular}
\end{table}

\begin{figure}[hhh]
\begin{center}
\scalebox{0.8}{\includegraphics{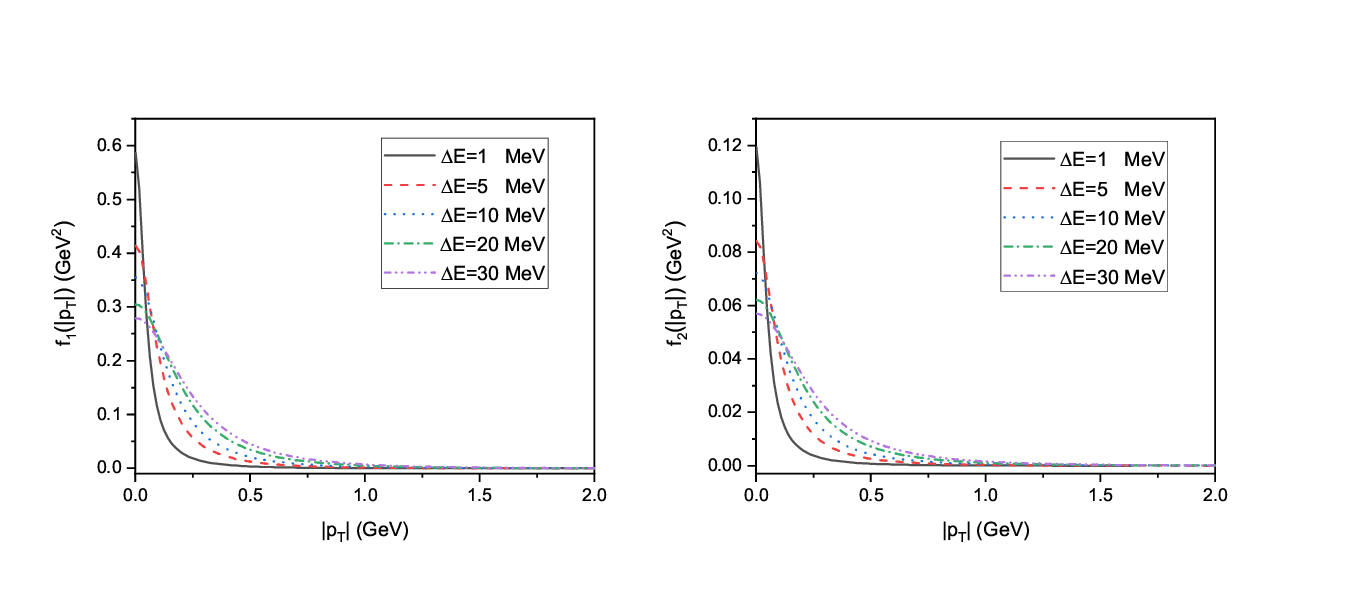}}
\end{center}
\caption{The unnormalized wave function $f_1(|\mathbf{p}_T|)$ and
$f_2(|\mathbf{p}_T|)$ for $\Sigma_c\bar D^*$ system with $I(J^P)=\frac{1}{2}(\frac{1}{2}^-)$
}\label{wave}
\end{figure}

\section{conclusion and discussion}

With more and more exotic states (tetraquarks and pentaquarks) being experimentally discovered, the new hadron ``zoo" gradually emerges, do we face the same situation
Gell-Mann did half century ago? Even though the whole picture is still vague, the shape might be encouraging that the time of establishing a unique theory on the
exotic states which would be an extension of the Gell-Manns SU(2) quark model, is coming. Nowadays, the task of high energy physicists is to carefully study every newly discovered exotic state,
namely find out its inner structure and production/decay characteristics. The procedure will definitely enrich our understanding. Our present work is just along the route forward.

Within the B-S framework we explore the possible bound states which are
composed of a baryon ($\Lambda_c$, $\Sigma_c$, $\Xi_c$ or $\Xi'_c$) and $\bar D^*$. In our work the orbital angular momentum between two constituents is $0$ ($S$-wave) so the total spin and parity is
$\frac{1}{2}^-$ or $\frac{3}{2}^-$. By solving the corresponding B-S equations of $\Lambda_c \bar D^*$, $\Sigma_c \bar D^*$, $\Xi_c \bar D^*$ and $\Xi'_c \bar D^*$ we  get possible information of bound states. If the B-S equation for
a supposed molecular structure has a solution and the parameters are reasonable, we would
conclude that the concerned pentaquark could exist in the nature,
oppositely, no-solution or the unreasonable parameter(s) means the supposed pentaquark cannot
appear as a resonance or the molecular state is not an appropriate
postulate. The solution can apply as a criterion for the
structures of the pentaquark states which have already been or
will be experimentally measured. In this work, the two
constituents interact by exchanging some light mesons. For the
$\Lambda_c\bar D^*$  system only $\omega$ and $\sigma$ are the
exchanged mediate mesons, while for the $\Sigma_c\bar D^*$ system
$\pi$, $\eta$ $\sigma$, $\rho$ and $\omega$ contribute. Similarly,  for $\Xi_c \bar D^*$ system only $\rho$, $\omega$ and  $\sigma$ can be exchanged, however for $\Xi'_c \bar D^*$ system $\pi$, $\eta$, $\sigma$, $\rho$ and $\omega$ can be
mediators. The chiral
interaction determines if those molecular states can be formed.

The key strategy we adopt in this work is that since two constituents are heavy hadrons, the B-S wave function
contains two scalar functions $f_1(|\mathbf{p_T}|)$ and
$f_2(|\mathbf{p_T}|)$ which should be solved numerically. After some manipulations the B-S equation turn into two coupled integral equations.
Discretizing the two coupled integral equations, we simplify them
into two algebraic equations about $f_1(Q_i)$
and $f_2(Q_i)$ $(i=1,2,...,n)$.
As $|\mathbf{p_T}|$ can take $n$ discrete values the two
coupled equations are converted into $2n$ algebraic equations which constitute a homogeneous linear equation group and can
be easily solved numerically in terms of available softwares.
When
all known parameters are input there still is one undetermined
parameter $\Lambda$. Our strategy is inputting binding energies
within a range and then fixing $\Lambda$ by solving the matrix
equation. If $\Lambda$ is located in a reasonable range one can
expect the bound state to exist. We find the B-S equation of  the
state $\Lambda_c \bar D^*$ system has no solution for $\Lambda$ when
the binding energy takes experimentally allowed values. For the
$\Sigma_c \bar D^*$ system there are four states with $I(J^P)=\frac{1}{2}(\frac{1}{2}^-)$, $\frac{1}{2}(\frac{3}{2}^-)$, $\frac{3}{2}(\frac{1}{2}^-)$ and $\frac{3}{2}(\frac{3}{2}^-)$.  We find the equation
for $I(J^P)=\frac{1}{2}(\frac{1}{2}^-)$ has a solution for $\Lambda$
falling into the reasonable range. The result supports that $P_c(4440)$ maybe
is a molecular state of $\Sigma_c \bar D^*$ with $I(J^P)=\frac{1}{2}(\frac{1}{2}^-)$.
Some authors suggest that $P_c(4457)$ should be
a bound state $\Sigma_c$ and $\bar
D^*$ with $I(J^P)=\frac{1}{2}(\frac{3}{2}^-)$. Our calculation indicates even for $\Delta E=1$ the value of $\Lambda$ is 1.927 GeV which deviates 1 GeV too much so the systme of $\Sigma_c\bar
D^*$ with  $I(J^P)=\frac{1}{2}(\frac{3}{2}^-)$ should not a resonant.

We also study the possible $\Xi_c\bar
D^*$ and $\Xi'_c\bar
D^*$ bound states. When no other effect is included the system $\Xi'_c\bar
D^*$ with $I(J^P)=0(\frac{1}{2}^-)$ can be a typical bound state, $\Xi_c\bar
D^*$ with $I(J^P)=0(\frac{1}{2}^-)$ and  $0(\frac{3}{2}^-)$ may be two very weak bound states at most  but $\Xi'_c\bar
D^*$ with $I(J^P)=0(\frac{3}{2}^-)$ shouldn't exist.
However after the effect of the mixing of $\Xi_c$ and $\Xi'_c$ is included the situation changed apparently.
$\Xi_c\bar
D^*$ with $I(J^P)=0(\frac{1}{2}^-)$ and $\Xi'_c\bar
D^*$ with $I(J^P)=0(\frac{1}{2}^-)$  may be two typical  bound states but $\Xi_c\bar
D^*$ with $I(J^P)=0(\frac{3}{2}^-)$  may be a weak bound state.

It is noted, we ignore the coupled
channel effects in the Bethe-Salpeter. If the coupled channel
interaction is taken
into account, just as
the authors of Refs. \cite{Wang:2022mxy,Shen:2017ayv} did, the value of $\Lambda$ may  change a little bit.

Within the B-S framework, we investigate the
bound state of a charmed baryon and $\bar D^*$. We pay a special attention to the systems $\Lambda_c \bar D^*$, $\Sigma_c \bar D^*$, $\Xi_c \bar D^*$ and $\Xi'_c \bar D^*$
because experimentally well measured $P_c(4440)$, $P_c(4457)$ and $P_{cs}(4459)$ may be related to them. From that
study, we have accumulated valuable knowledge on probable
molecular structure of pentaquarks which can be applied to the
future research.  Definitely, the discovery of pentaquarks opens a
window for understanding the quark model established by Gell-Mann
and several other predecessors. Deeper study on their structures
and concerned effective interaction which binds the ingredients to
form a molecule would greatly enrich our theoretical aspect.

\section*{Acknowledgement}
This work is supported by the National Natural Science Foundation
of China (NNSFC) under the contract No. 12075167, 11975165, 12235018,
 12075125, 12035009 and 11735010.

\appendix
\section{The effective interactions}

The effective interactions $\mathcal{BBM}$ can be found
in \cite{Ronchen:2012eg,Shen:2016tzq,He:2017aps}
\begin{eqnarray}
&&\mathcal{L}_{\mathcal{B}_{\bar 3}\mathcal{B}_{\bar 3}\sigma}=l_\mathcal{B}<\bar\mathcal{B}_{\bar 3}\sigma\mathcal{B}_{\bar 3}>\\
&&\mathcal{L}_{\mathcal{B}_{\bar 3}\mathcal{B}_{\bar 3}\mathcal{V}}=\frac{\beta_{\mathcal{B}}g_{\mathcal{V}}}{\sqrt{2}}<\bar\mathcal{B}_{\bar 3}v\cdot \mathcal{V}\mathcal{B}_{\bar 3}>,
\\&&\mathcal{L}_{\mathcal{B}_6\mathcal{B}_6\sigma}=-l_\mathcal{S}<\bar\mathcal{B}_6\sigma\mathcal{B}_6>,
\\&&\mathcal{L}_{\mathcal{B}_6\mathcal{B}_6\mathcal{P}}=i\frac{g_1}{2f_\pi}\varepsilon^{\mu\nu\alpha\beta}v_\beta l_\mathcal{S}<\bar\mathcal{B}_6\gamma_\mu\gamma_\alpha\partial_\nu \mathcal{P}\mathcal{B}_6>,
\\&&\mathcal{L}_{\mathcal{B}_6\mathcal{B}_6\mathcal{V}}=-\frac{\beta_{\mathcal{S}}g_{\mathcal{V}}}{\sqrt{2}}<\bar\mathcal{B}_6v\cdot \mathcal{V}\mathcal{B}_6>
-i\frac{\lambda_{\mathcal{S}}g_{\mathcal{V}}}{3\sqrt{2}}<\bar\mathcal{B}_6\gamma_\mu\gamma_\nu(\partial^\mu \mathcal{V}^\nu-\partial^\nu \mathcal{V}^\mu)\mathcal{B}_6>\end{eqnarray}
where
$
\mathcal{B}_{\bar {\bar 3}}=\left(\begin{array}{ccc}
        0&\Lambda_c^+ &\Xi_c^+ \\
         -\Lambda_c^+ & 0&\Xi_c^0\\
       - \Xi_c^+ & \Xi_c^0 &0
      \end{array}\right)$,
      $
\mathcal{B}_6=\left(\begin{array}{ccc}
        \Sigma_{c}^{+}& \frac{\Sigma_{c}^{+}}{\sqrt{2}}&\frac{\Xi_c^{'+}}{\sqrt{2}} \\
         \frac{\Sigma_{c}^{+}}{\sqrt{2}}& \Sigma_{c}^{0}&\frac{\Xi_c^{'0}}{\sqrt{2}} \\
  \frac{\Xi_c^{'+}}{\sqrt{2}} & \frac{\Xi_c^{'0}}{\sqrt{2}} &\Omega_c^0
      \end{array}\right)$,
$
\mathcal{P}=\left(\begin{array}{ccc}
        \frac{\pi^0}{\sqrt{2}}+\frac{\eta}{\sqrt{6}} &\pi^+ &K^+ \\
         \pi^- & -\frac{\pi^0}{\sqrt{2}}+\frac{\eta}{\sqrt{6}}&K^0\\
         K^-& \bar{K^0} & -\sqrt{\frac{2}{3}}\eta
      \end{array}\right)$
       and $
\mathcal{V}=\left(\begin{array}{ccc}
        \frac{\rho^0}{\sqrt{2}}+\frac{\omega}{\sqrt{2}} &\rho^+ &K^{*+} \\
         \rho^- & -\frac{\rho^0}{\sqrt{2}}+\frac{\omega}{\sqrt{2}}&K^{*0}\\
         K^{*-}& \bar{K^{*0}} & \phi
      \end{array}\right)$, respectively.

  The effective interactions $\bar D\bar D \mathcal{M}$ can be found
in \cite{Colangelo:2005gb,Colangelo:2012xi,Ding:2008gr}
\begin{eqnarray}
&&\mathcal{L}_{_{\bar D^*\bar D^*\sigma}}=g_{_{\bar D^*\bar D^*\sigma}}(\bar D^{*\mu}_{b}
\bar D^{*\dag}_{a\mu}\sigma),
\\&&\mathcal{L}_{_{\bar D^*\bar D^*\mathcal{P}}}=g_{_{\bar D^*\bar D^*\mathcal{P}}}(\bar D^{*\mu}_{b}\stackrel{\leftrightarrow}{\partial}^{\beta}
\bar D^{*\alpha\dag}_{a})(\partial^\nu
\mathcal{P})_{ab}\varepsilon_{\nu\mu\alpha\beta},
\\&&\mathcal{L}_{_{\bar D^*\bar D^*\mathcal{V}}}=ig_{_{\bar D^*\bar D^*\mathcal{V}}}(\bar D^{*\nu}_{b}\stackrel{\leftrightarrow}{\partial}_{\mu}
\bar D^{*\dag}_{a\nu})(
\mathcal{V})_{ab}^\mu+ig'_{_{\bar D^*\bar D^*\mathcal{V}}}\bar D^{*\mu}_{b}
\bar D^{*\nu\dag}_{a}(
\partial_\mu\mathcal{V}_\nu-\partial_\nu\mathcal{V}_\mu)_{ab}
\end{eqnarray}
where $a$ and $b$ represent index of SU(3) flavor group for three light quarks.
In the flavor $SU(3)$ symmetry and heavy quark limit, the above coupling constants
are given by$g_{_{D^*D^*\sigma}}=2g_\sigma M_{D^*},$
$g_{_{D^*D^*P}}=\frac{2g}{f_\pi},$ $ g_{_{\bar D^*\bar D^*V}}=\beta
g_V/\sqrt{2},\,\, g'_{_{\bar D^*\bar D^*V}}=2\sqrt{2}\lambda g_V M_{D^*}$
with $g_\sigma=-0.76$ \cite{Bardeen:2003kt}, $f_\pi=132$ MeV \cite{Colangelo:2005gb},
 $\beta=0.9$,
$g_V=5.9$ \cite{Falk:1992cx} and $\lambda =0.56$ GeV$^{-1}$ \cite{Chen:2019asm}.

\section{The coupled equation of $f_1(|\mathbf{p}_T|)$ and
$f_2(|\mathbf{p}_T|)$ after integrating over $p_l$ and some formulas for azimuthal integration}

The coupled equation of $f_1(|\mathbf{p}_T|)$ and
$f_2(|\mathbf{p}_T|)$ after integrating over $p_l$  are
\begin{eqnarray} \label{couple equation12}
&&f_1(|\mathbf{p}_T|)=\int\frac{d^3\mathbf{q}_T}{(2\pi)^3}\frac{1}
{P_{\mathcal{R}1}}\{
[\sum_\mathcal{P}\frac{C_{{I\mathcal{P}}}g_{_{1\mathcal{P}}}g_{_{2\mathcal{P}}}F^2(k,m_\mathcal{P})K^\mathcal{P}_{1}}{-(\mathbf{p}_T-\mathbf{q}_T)^2-m_\mathcal{P}^2}+
\frac{C_{{I\mathcal{S}}}g_{_{1\mathcal{S}}}g_{_{2\mathcal{S}}}F^2(k,m_\mathcal{S})K^\mathcal{S}_{1}}{-(\mathbf{p}_T-\mathbf{q}_T)^2-m_\mathcal{S}^2}\nonumber\\&&+\sum_\mathcal{V}
\frac{-C_{{I\mathcal{V}}}F^2(k,m_\mathcal{V})(g_{_{1\mathcal{V}}}g_{_{2\mathcal{V}}}K^{\mathcal{V},a}_{1}+g_{_{1\mathcal{V}}}g'_{_{2\mathcal{V}}}K^{\mathcal{V},b}_{1}
+g'_{_{1\mathcal{V}}}g_{_{2\mathcal{V}}}K^{\mathcal{V},c}_{1}+g'_{_{1\mathcal{V}}}g'_{_{2\mathcal{V}}}K^{\mathcal{V},d}_{1})}{-(\mathbf{p}_T-\mathbf{q}_T)^2-m_\mathcal{V}^2}
]_{p_l=-\eta_1M-\omega_1}\}\nonumber\\&&+
\int\frac{d^3\mathbf{q}_T}{(2\pi)^3}\frac{1}
{P_{\mathcal{R}2}}\{[\sum_\mathcal{P}\frac{C_{{I\mathcal{P}}}g_{_{1\mathcal{P}}}g_{_{2\mathcal{P}}}F^2(k,m_\mathcal{P})K^\mathcal{P}_{1}}{-(\mathbf{p}_T-\mathbf{q}_T)^2-m_\mathcal{P}^2}
+\frac{C_{{I\mathcal{S}}}g_{_{1\mathcal{S}}}g_{_{2\mathcal{S}}}F^2(k,m_\mathcal{S})K^\mathcal{S}_{1}}{-(\mathbf{p}_T-\mathbf{q}_T)^2-m_\mathcal{S}^2}
\nonumber\\&&-\sum_\mathcal{V}
\frac{C_{{I\mathcal{V}}}F^2(k,m_\mathcal{V})(g_{_{1\mathcal{V}}}g_{_{2\mathcal{V}}}K^{\mathcal{V},a}_{1}+g_{_{1\mathcal{V}}}g'_{_{2\mathcal{V}}}K^{\mathcal{V},b}_{1}
+g'_{_{1\mathcal{V}}}g_{_{2\mathcal{V}}}K^{\mathcal{V},c}_{1}+g'_{_{1\mathcal{V}}}g'_{_{2\mathcal{V}}}K^{\mathcal{V},d}_{1})}{-(\mathbf{p}_T-\mathbf{q}_T)^2-m_\mathcal{V}^2}]_{p_l=\eta_2M-\omega_2}\}.
\end{eqnarray}

\begin{eqnarray} \label{couple equation22}
&&f_2(|\mathbf{p}_T|)=\int\frac{d^3\mathbf{q}_T}{(2\pi)^3}\frac{1}
{P_{\mathcal{R}1}}\{
[\sum_\mathcal{P}\frac{C_{{I\mathcal{P}}}g_{_{1\mathcal{P}}}g_{_{2\mathcal{P}}}F^2(k,m_\mathcal{P})K^\mathcal{P}_{2}}{-(\mathbf{p}_T-\mathbf{q}_T)^2-m_\mathcal{P}^2}
+\frac{C_{{I\mathcal{S}}}g_{_{1\mathcal{S}}}g_{_{2\mathcal{S}}}F^2(k,m_\mathcal{S})K^\mathcal{S}_{2}}{-(\mathbf{p}_T-\mathbf{q}_T)^2-m_\mathcal{S}^2}\nonumber\\&&+\sum_\mathcal{V}
\frac{-C_{{I\mathcal{V}}}F^2(k,m_\mathcal{V})(g_{_{1\mathcal{V}}}g_{_{2\mathcal{V}}}K^{\mathcal{V},a}_{2}+g_{_{1\mathcal{V}}}g'_{_{2\mathcal{V}}}K^{\mathcal{V},b}_{2}
+g'_{_{1\mathcal{V}}}g_{_{2\mathcal{V}}}K^{\mathcal{V},c}_{2}+g'_{_{1\mathcal{V}}}g'_{_{2\mathcal{V}}}K^{\mathcal{V},d}_{2})}{-(\mathbf{p}_T-\mathbf{q}_T)^2-m_\mathcal{V}^2}
]_{p_l=-\eta_1M-\omega_1}\}\nonumber\\&&+
\int\frac{d^3\mathbf{q}_T}{(2\pi)^3}\frac{1}
{P_{\mathcal{R}2}}\{[\sum_\mathcal{P}\frac{C_{{I\mathcal{P}}}g_{_{1\mathcal{P}}}g_{_{2\mathcal{P}}}F^2(k,m_\mathcal{P})K^\mathcal{P}_{2}}{-(\mathbf{p}_T-\mathbf{q}_T)^2-m_\mathcal{P}^2}
+\frac{C_{{I\mathcal{S}}}g_{_{1\mathcal{S}}}g_{_{2\mathcal{S}}}F^2(k,m_\mathcal{S})K^\mathcal{S}_{2}}{-(\mathbf{p}_T-\mathbf{q}_T)^2-m_\mathcal{S}^2}\nonumber\\&&-\sum_\mathcal{V}
\frac{C_{{I\mathcal{V}}}F^2(k,m_\mathcal{V})(g_{_{1\mathcal{V}}}g_{_{2\mathcal{V}}}K^{\mathcal{V},a}_{2}+g_{_{1\mathcal{V}}}g'_{_{2\mathcal{V}}}K^{\mathcal{V},b}_{2}
+g'_{_{1\mathcal{V}}}g_{_{2\mathcal{V}}}K^{\mathcal{V},c}_{2}+g'_{_{1\mathcal{V}}}g'_{_{2\mathcal{V}}}K^{\mathcal{V},d}_{2})}{-(\mathbf{p}_T-\mathbf{q}_T)^2-m_\mathcal{V}^2}]_{p_l=\eta_2M-\omega_2}\},
\end{eqnarray}
with $P_{\mathcal{R}1}=2\omega_1(M+\omega_1+\omega_2)(M+\omega_1-\omega_2)$ and $P_{\mathcal{R}2}=2\omega_2(M+\omega_1-\omega_2)(M-\omega_1-\omega_2)$.

Since $d^3\mathbf{q}_T=\mathbf{q}_T^2{\rm
sin}(\theta)d|\mathbf{q}_T|d\theta d\phi$ and $\mathbf{p}_T\cdot
\mathbf{q}_T=|\mathbf{p}_T||\mathbf{q}_T|{\rm cos}(\theta)$ one
can carry out the azimuthal integration for Eqs. (\ref{couple
equation12}) and (\ref{couple equation22}) analytically. Some
useful integrations are defined as follow

\begin{eqnarray} \label{azimuthal1}
&&J_0\equiv\int_0^\pi{\rm sin}(\theta)d\theta
\frac{1}{-(\mathbf{p}_T-\mathbf{q}_T)^2-m_\mathcal{M}^2}[\frac{\Lambda^2-m^2_\mathcal{M}}{\Lambda^2-{(\mathbf{p}_T-\mathbf{q}_T)^2}}]^2
\nonumber\\&&=\int_0^\pi \frac{{\rm
sin}(\theta)d\theta}{-[\mathbf{p}_T^2+\mathbf{q}_T^2-2|\mathbf{p}_T||\mathbf{q}_T|{\rm
cos}(\theta)]-m_\mathcal{M}^2}\{\frac{\Lambda^2-m^2_\mathcal{M}}{\Lambda^2-{[\mathbf{p}_T^2+\mathbf{q}_T^2-2|\mathbf{p}_T||\mathbf{q}_T|{\rm
cos}(\theta)]}}\}^2\nonumber\\&&=-\frac{2(m_\mathcal{M}^2-\Lambda^2)}{[(|\mathbf{p}_T|-|\mathbf{q}_T|)^2+\Lambda^2][(|\mathbf{p}_T|+|\mathbf{q}_T|)^2
+\Lambda^2]}\nonumber\\&&+\frac{1}{2|\mathbf{p}_T||\mathbf{q}_T|}\{{\rm
Ln}[\frac{(|\mathbf{p}_T|+|\mathbf{q}_T|)^2+\Lambda^2}{(|\mathbf{p}_T|-|\mathbf{q}_T|)^2+\Lambda^2}]-{\rm
Ln}[\frac{(|\mathbf{p}_T|+
|\mathbf{q}_T|)^2+m_\mathcal{M}^2}{(|\mathbf{p}_T|-|\mathbf{q}_T|)^2+m_\mathcal{M}^2}]\},
\end{eqnarray}

\begin{eqnarray} \label{azimuthal2}
&&J_1\equiv\int_0^\pi {\rm
sin}(\theta)d\theta\frac{\mathbf{p}_T\cdot
\mathbf{q}_T}{-(\mathbf{p}_T-\mathbf{q}_T)^2-m_\mathcal{M}^2}[\frac{\Lambda^2-m^2_\mathcal{M}}{\Lambda^2-{(\mathbf{p}_T-\mathbf{q}_T)^2}}]^2
\nonumber\\&&=\int_0^\pi \frac{|\mathbf{p}_T||\mathbf{q}_T|{\rm
cos}(\theta){\rm
sin}(\theta)d\theta}{-[\mathbf{p}_T^2+\mathbf{q}_T^2-2|\mathbf{p}_T||\mathbf{q}_T|{\rm
cos}(\theta)]-m_\mathcal{M}^2}\{\frac{\Lambda^2-m^2_\mathcal{M}}{\Lambda^2-{[\mathbf{p}_T^2+\mathbf{q}_T^2-2|\mathbf{p}_T||\mathbf{q}_T|{\rm
cos}(\theta)]}}\}^2\nonumber\\&&=-\frac{(m_\mathcal{M}^2-\Lambda^2)(|\mathbf{p}_T|^2+|\mathbf{q}_T|^2
+\Lambda^2)}{[(|\mathbf{p}_T|-|\mathbf{q}_T|)^2+\Lambda^2][(|\mathbf{p}_T|+|\mathbf{q}_T|)^2
+\Lambda^2]}\nonumber\\&&+\frac{(|\mathbf{p}_T|^2+|\mathbf{q}_T|^2
+m_\mathcal{M}^2)}{4|\mathbf{p}_T||\mathbf{q}_T|}\{Ln[\frac{(|\mathbf{p}_T|+|\mathbf{q}_T|)^2+\Lambda^2}{(|\mathbf{p}_T|-|\mathbf{q}_T|)^2+\Lambda^2}]-Ln[\frac{(|\mathbf{p}_T|+
|\mathbf{q}_T|)^2+m_\mathcal{M}^2}{(|\mathbf{p}_T|-|\mathbf{q}_T|)^2+m_\mathcal{M}^2}]\},
\end{eqnarray}

\begin{eqnarray} \label{azimuthal3}
&&J_2\equiv\int_0^\pi {\rm
sin}(\theta)d\theta\frac{(\mathbf{p}_T\cdot
\mathbf{q}_T)^2}{-(\mathbf{p}_T-\mathbf{q}_T)^2-m_\mathcal{M}^2}[\frac{\Lambda^2-m^2_\mathcal{M}}{\Lambda^2-{(\mathbf{p}_T-\mathbf{q}_T)^2}}]^2
\nonumber\\&&=\int_0^\pi
\frac{|\mathbf{p}_T|^2|\mathbf{q}_T|^2{\rm cos}^2(\theta){\rm
sin}(\theta)d\theta}{-[\mathbf{p}_T^2+\mathbf{q}_T^2-2|\mathbf{p}_T||\mathbf{q}_T|{\rm
cos}(\theta)]-m_\mathcal{M}^2}\{\frac{\Lambda^2-m^2_\mathcal{M}}{\Lambda^2-{[\mathbf{p}_T^2+\mathbf{q}_T^2-2|\mathbf{p}_T||\mathbf{q}_T|{\rm
cos}(\theta)]}}\}^2\nonumber\\&&=-\frac{(m_\mathcal{M}^2-\Lambda^2)(|\mathbf{p}_T|^2+|\mathbf{q}_T|^2
+\Lambda^2)^2}{2[(|\mathbf{p}_T|-|\mathbf{q}_T|)^2+\Lambda^2][(|\mathbf{p}_T|+|\mathbf{q}_T|)^2
+\Lambda^2]}\nonumber\\&&+\frac{1}{8|\mathbf{p}_T||\mathbf{q}_T|}\{(|\mathbf{p}_T|^2+|\mathbf{q}_T|^2
+2m_\mathcal{M}^2-\Lambda^2)(|\mathbf{p}_T|^2+|\mathbf{q}_T|^2
+\Lambda^2){\rm
Ln}[\frac{(|\mathbf{p}_T|+|\mathbf{q}_T|)^2+\Lambda^2}{(|\mathbf{p}_T|-|\mathbf{q}_T|)^2+\Lambda^2}]\nonumber\\&&-(|\mathbf{p}_T|^2+|\mathbf{q}_T|^2
+m_\mathcal{M}^2)^2{\rm Ln}[\frac{(|\mathbf{p}_T|+
|\mathbf{q}_T|)^2+m_\mathcal{M}^2}{(|\mathbf{p}_T|-|\mathbf{q}_T|)^2+m_\mathcal{M}^2}]\}.
\end{eqnarray}

The detail expressions of $A_{11}(\mathbf{p_T},\mathbf{q_T})$, $A_{12}(\mathbf{p_T},\mathbf{q_T})$, $A_{21}(\mathbf{p_T},\mathbf{q_T})$ and $A_{22}(\mathbf{p_T},\mathbf{q_T})$ are
\begin{eqnarray} \label{couple equation13}
&&A_{11}(\mathbf{p_T},\mathbf{q_T})=\frac{\mathbf{q}_T^2}{(2\pi)^2}\frac{1}
{P_{\mathcal{R}1}}[\sum_\mathcal{P} C_{{I\mathcal{P}}}C_{J\mathcal{P}}g_{_{1\mathcal{P}}}g_{_{2\mathcal{P}}}A_{11}^{\mathcal{P}}+
C_{{I\mathcal{S}}}C_{J\mathcal{S}}g_{_{1\mathcal{S}}}g_{_{2\mathcal{S}}}A_{11}^{\mathcal{S}}\nonumber\\&&-
\sum_\mathcal{V} C_{{I\mathcal{V}}}(C_{Ja}g_{_{1\mathcal{V}}}g_{_{2\mathcal{V}}}A_{11}^{\mathcal{V},a}+C_{Jb}g_{_{1\mathcal{V}}}g'_{_{2\mathcal{V}}}A_{11}^{\mathcal{V},b}+
C_{Jc}g'_{_{1\mathcal{V}}}g_{_{2\mathcal{V}}}A_{11}^{\mathcal{V},c}+C_{Jd}g'_{_{1\mathcal{V}}}g'_{_{2\mathcal{V}}}A_{11}^{\mathcal{V},d})]+\nonumber\\&&
\frac{\mathbf{q}_T^2}{(2\pi)^2}\frac{1
}
{P_{\mathcal{R}2}}[\sum_\mathcal{P} C_{{I\mathcal{P}}}C_{J\mathcal{P}}g_{_{1\mathcal{P}}}g_{_{2\mathcal{P}}}A_{11}^{'\mathcal{P}}
+ C_{{I\mathcal{S}}}C_{J\mathcal{S}}g_{_{1\mathcal{S}}}g_{_{2\mathcal{S}}}A_{11}^{'\mathcal{S}}\nonumber\\&&-
\sum_\mathcal{V} C_{{I\mathcal{V}}}(C_{Ja}g_{_{1\mathcal{V}}}g_{_{2\mathcal{V}}}A_{11}^{'\mathcal{V},a}+C_{Jb}g_{_{1\mathcal{V}}}g'_{_{2\mathcal{V}}}A_{11}^{'\mathcal{V},b}+
C_{Jc}g'_{_{1\mathcal{V}}}g_{_{2\mathcal{V}}}A_{11}^{'\mathcal{V},c}+C_{Jd}g'_{_{1\mathcal{V}}}g'_{_{2\mathcal{V}}}A_{11}^{'\mathcal{V},d})],
\end{eqnarray}

\begin{eqnarray} \label{couple equation14}
&&A_{12}(\mathbf{p_T},\mathbf{q_T})=\frac{\mathbf{q}_T^2}{(2\pi)^2}\frac{1}
{P_{\mathcal{R}1}}[\sum_\mathcal{P} C_{{I\mathcal{P}}}C_{J\mathcal{P}}g_{_{1\mathcal{P}}}g_{_{2\mathcal{P}}}A_{12}^{\mathcal{P}}+
C_{{I\mathcal{S}}}C_{J\mathcal{S}}g_{_{1\mathcal{S}}}g_{_{2\mathcal{S}}}A_{12}^{\mathcal{S}}\nonumber\\&&-
\sum_\mathcal{V} C_{{I\mathcal{V}}}(C_{Ja}g_{_{1\mathcal{V}}}g_{_{2\mathcal{V}}}A_{12}^{\mathcal{V},a}+C_{Jb}g_{_{1\mathcal{V}}}g'_{_{2\mathcal{V}}}A_{12}^{\mathcal{V},b}+
C_{Jc}g'_{_{1\mathcal{V}}}g_{_{2\mathcal{V}}}A_{12}^{\mathcal{V},c}+C_{Jd}g'_{_{1\mathcal{V}}}g'_{_{2\mathcal{V}}}A_{12}^{\mathcal{V},d})]+\nonumber\\&&
\frac{\mathbf{q}_T^2}{(2\pi)^2}\frac{1
}
{P_{\mathcal{R}2}}[\sum_\mathcal{P} C_{{I\mathcal{P}}}C_{J\mathcal{P}}g_{_{1\mathcal{P}}}g_{_{2\mathcal{P}}}A_{12}^{'\mathcal{P}}
+C_{J\mathcal{S}}g_{_{1\mathcal{S}}}g_{_{2\mathcal{S}}}A_{12}^{'\mathcal{S}}\nonumber\\&&-
\sum_\mathcal{V} C_{{I\mathcal{V}}}(C_{Ja}g_{_{1\mathcal{V}}}g_{_{2\mathcal{V}}}A_{12}^{'\mathcal{V},a}+C_{Jb}g_{_{1\mathcal{V}}}g'_{_{2\mathcal{V}}}A_{12}^{'\mathcal{V},b}+
C_{Jc}g'_{_{1\mathcal{V}}}g_{_{2\mathcal{V}}}A_{12}^{'\mathcal{V},c}+C_{Jd}g'_{_{1\mathcal{V}}}g'_{_{2\mathcal{V}}}A_{12}^{'\mathcal{V},d})],
\end{eqnarray}

\begin{eqnarray} \label{couple equation23}
&&A_{21}(\mathbf{p_T},\mathbf{q_T})=\frac{\mathbf{q}_T^2}{(2\pi)^2}\frac{1}
{P_{\mathcal{R}1}}[\sum_\mathcal{P} C_{{I\mathcal{P}}}C_{J\mathcal{P}}g_{_{1\mathcal{P}}}g_{_{2\mathcal{P}}}A_{21}^{\mathcal{P}}+
C_{{I\mathcal{S}}}C_{J\mathcal{S}}g_{_{1\mathcal{S}}}g_{_{2\mathcal{S}}}A_{21}^{\mathcal{S}}\nonumber\\&&-\sum_\mathcal{V} C_{{I\mathcal{V}}}(C_{Ja}g_{_{1\mathcal{V}}}g_{_{2\mathcal{V}}}A_{21}^{\mathcal{V},a}+C_{Jb}g_{_{1\mathcal{V}}}g'_{_{2\mathcal{V}}}A_{21}^{\mathcal{V},b}+
C_{Jc}g'_{_{1\mathcal{V}}}g_{_{2\mathcal{V}}}A_{21}^{\mathcal{V},c}+C_{Jd}g'_{_{1\mathcal{V}}}g'_{_{2\mathcal{V}}}A_{21}^{\mathcal{V},d})]+\nonumber\\&&
\frac{\mathbf{q}_T^2}{(2\pi)^2}\frac{1
}
{P_{\mathcal{R}2}}[\sum_\mathcal{P} C_{{I\mathcal{P}}}C_{J\mathcal{P}}g_{_{1\mathcal{P}}}g_{_{2\mathcal{P}}}A_{21}^{'\mathcal{P}}
+C_{{I\mathcal{S}}}C_{J\mathcal{S}}g_{_{1\mathcal{S}}}g_{_{2\mathcal{S}}}A_{21}^{'\mathcal{S}}\nonumber\\&&-
\sum_\mathcal{V} C_{{I\mathcal{V}}}(C_{Ja}g_{_{1\mathcal{V}}}g_{_{2\mathcal{V}}}A_{21}^{'\mathcal{V},a}+C_{Jb}g_{_{1\mathcal{V}}}g'_{_{2\mathcal{V}}}A_{21}^{'\mathcal{V},b}+
C_{Jc}g'_{_{1\mathcal{V}}}g_{_{2\mathcal{V}}}A_{21}^{'\mathcal{V},c}+C_{Jd}g'_{_{1\mathcal{V}}}g'_{_{2\mathcal{V}}}A_{21}^{'\mathcal{V},d})],
\end{eqnarray}

\begin{eqnarray} \label{couple equation24}
&&A_{22}(\mathbf{p_T},\mathbf{q_T})=\frac{\mathbf{q}_T^2}{(2\pi)^2}\frac{1}
{P_{\mathcal{R}1}}[\sum_\mathcal{P} C_{{I\mathcal{P}}}C_{J\mathcal{P}}g_{_{1\mathcal{P}}}g_{_{2\mathcal{P}}}A_{22}^{\mathcal{P}}+
C_{{I\mathcal{S}}}C_{J\mathcal{S}}g_{_{1\mathcal{S}}}g_{_{2\mathcal{S}}}A_{22}^{\mathcal{S}}\nonumber\\&&-\sum_\mathcal{V} C_{{I\mathcal{V}}}(C_{Ja}g_{_{1\mathcal{V}}}g_{_{2\mathcal{V}}}A_{22}^{\mathcal{V},a}+C_{Jb}g_{_{1\mathcal{V}}}g'_{_{2\mathcal{V}}}A_{22}^{\mathcal{V},b}+
C_{Jc}g'_{_{1\mathcal{V}}}g_{_{2\mathcal{V}}}A_{22}^{\mathcal{V},c}+C_{Jd}g'_{_{1\mathcal{V}}}g'_{_{2\mathcal{V}}}A_{22}^{\mathcal{V},d})]+\nonumber\\&&
\frac{\mathbf{q}_T^2}{(2\pi)^2}\frac{1
}
{P_{\mathcal{R}2}}[\sum_\mathcal{P} C_{{I\mathcal{P}}}C_{J\mathcal{P}}g_{_{1\mathcal{P}}}g_{_{2\mathcal{P}}}A_{22}^{'\mathcal{P}}
+C_{{I\mathcal{S}}}C_{J\mathcal{S}}g_{_{1\mathcal{S}}}g_{_{2\mathcal{S}}}A_{22}^{'\mathcal{S}}\nonumber\\&&-
\sum_\mathcal{V} C_{{I\mathcal{V}}}(C_{Ja}g_{_{1\mathcal{V}}}g_{_{2\mathcal{V}}}A_{22}^{'\mathcal{V},a}+C_{Jb}g_{_{1\mathcal{V}}}g'_{_{2\mathcal{V}}}A_{22}^{'\mathcal{V},b}+
C_{Jc}g'_{_{1\mathcal{V}}}g_{_{2\mathcal{V}}}A_{22}^{'\mathcal{V},c}+C_{Jd}g'_{_{1\mathcal{V}}}g'_{_{2\mathcal{V}}}A_{22}^{'\mathcal{V},d})].
\end{eqnarray}
with
\begin{eqnarray}
&&A_{11}^{\mathcal{S}}= (m_1-\omega_1)J_0,\nonumber\\
&&A_{11}^{'\mathcal{S}}=(M+m_1-\omega_2)J_0,\nonumber\\
&&A_{11}^{\mathcal{P}}=\frac{16}{3}J_0 (M+{\omega_1}) \left\{{\mathbf{p}_T}^2[{\omega_1} (m_1-{\omega_1})+{\mathbf{q}_T}^2]+{\mathbf{q}_T}^2\omega_1 (m_1-{\omega_1})\right\}\nonumber\\&&+\frac{32}{3}J_1
  \omega_1 (M+{\omega_1}) ({\omega_1}-m_1)-\frac{16}{3}J_2 (M+{\omega_1}),
\nonumber\\&&
A_{11}^{'\mathcal{P}}=\frac{16}{3}J_0\omega_2 \left\{-[(M-\omega_2)^2+(M-\omega_2)  m_1] ({\mathbf{p}_T}^2+ {\mathbf{q}_T}^2)+{\mathbf{p}_T}^2 {\mathbf{q}_T}^2\right\}\nonumber\\&&+\frac{32}{3}J_1\omega_2 [ M^2+
   M (m_1-\omega_2)+2\omega_2 ({\omega_2}-m_1)]-\frac{16}{3}J_2\omega_2,\nonumber\\&&
A_{11}^{\mathcal{V},a}=-J_0 (m_1-{\omega_1}) [-4 {m_\mathcal{V}}^2\omega_1 (M+{\omega_1})+{m_\mathcal{V}}^2 {\mathbf{p}_T}^2+{m_\mathcal{V}}^2
   {\mathbf{q}_T}^2+({\mathbf{p}_T}^2-{\mathbf{q}_T}^2)^2]/{{m_\mathcal{V}}^2}\nonumber\\&&-2J_1 (m_1-{\omega_1}),\nonumber\\&&
A_{11}^{'\mathcal{V},a}=-J_0 (M+m_1-{\omega_2}) [4 {m_\mathcal{V}}^2\omega_2 (M-{\omega_2})+{m_\mathcal{V}}^2 {\mathbf{p}_T}^2+{m_\mathcal{V}}^2
   {\mathbf{q}_T}^2+({\mathbf{p}_T}^2-{\mathbf{q}_T}^2)^2]/{{m_\mathcal{V}}^2}\nonumber\\&&-2J_1 (M+m_1-{\omega_2}),
\nonumber\\&&
A_{11}^{\mathcal{V},b}=A_{11}^{'\mathcal{V},b}=0,\nonumber\\&&
A_{11}^{\mathcal{V},c}=4J_0 {\mathbf{p}_T}^2 (M+{\omega_1})-4J_1 (M+{\omega_1}),\nonumber\\&&
A_{11}^{'\mathcal{V},c}=4J_0 {\mathbf{p}_T}^2\omega_2-4J_1\omega_2,\nonumber\\&&
A_{11}^{\mathcal{V},d}=\frac{8}{3}J_0 (m_1-{\omega_1}) (-{\mathbf{p}_T}^2-{\mathbf{q}_T}^2)+\frac{16}{3}J_1 (m_1-{\omega_1}),
\nonumber\\&&
A_{11}^{'\mathcal{V},d}=\frac{8}{3}J_0 (-{\mathbf{p}_T}^2-{\mathbf{q}_T}^2) (M+m_1-{\omega_2})+\frac{16}{3}J_1 (M+m_1-{\omega_2}),
\nonumber\\&&
\end{eqnarray}

\begin{eqnarray}
&&A_{12}^{\mathcal{S}}= -J_1,\nonumber\\
&&A_{12}^{'\mathcal{S}}=-J_1,\nonumber\\
&&A_{12}^{\mathcal{P}}=\frac{16}{3}[J_0 (M+{\omega_1}) (m_1 +\omega_1){\mathbf{p}_T}^2 {\mathbf{q}_T}^2-J_1 (M+{\omega_1}) ({\mathbf{p}_T}^2
  +{\mathbf{q}_T}^2)\omega_1+J_2 (M+{\omega_1}) ({\omega_1}-m_1)],
\nonumber\\&&
A_{12}^{'\mathcal{P}}=-\frac{16}{3}[J_0\omega_2 ( M-{\omega_2}-m_1){\mathbf{p}_T}^2 {\mathbf{q}_T}^2-J_1\omega_2 ({\mathbf{p}_T}^2
   +{\mathbf{q}_T}^2 )(M-{\omega_2})+J_2\omega_2 (M+m_1-{\omega_2})],\nonumber\\&&
A_{12}^{\mathcal{V},a}=J_1 [-4 {m_\mathcal{V}}^2\omega_1 (M+{\omega_1})+{m_\mathcal{V}}^2 {\mathbf{p}_T}^2+{m_\mathcal{V}}^2 {\mathbf{q}_T}^2+({\mathbf{p}_T}^2-{\mathbf{q}_T}^2)^2]/{m_\mathcal{V}}^2+2
  J_2,\nonumber\\&&
A_{12}^{'\mathcal{V},a}=J_1 [4 {m_\mathcal{V}}^2\omega_2 (M-{\omega_2})+{m_\mathcal{V}}^2 {\mathbf{p}_T}^2+{m_\mathcal{V}}^2 {\mathbf{q}_T}^2+({\mathbf{p}_T}^2-{\mathbf{q}_T}^2)^2]/{{m_\mathcal{V}}^2}+2
  J_2,\nonumber\\&&
A_{12}^{\mathcal{V},b}=A_{12}^{'\mathcal{V},b}=\frac{4}{3}[J_0 {\mathbf{p}_T}^2 {\mathbf{q}_T}^2-J_2],
\nonumber\\&&
A_{12}^{\mathcal{V},c}=4[J_0{\mathbf{q}_T}^2 [(M+{\omega_1}) (m_1-{\omega_1})+ {\mathbf{p}_T}^2]-J_1 (M
  +{\omega_1}) (m_1-{\omega_1})-J_2],
\nonumber\\&&
A_{12}^{'\mathcal{V},c}=4[J_0 {\mathbf{q}_T}^2 [{\omega_2} (M+m_1-{\omega_2})+{\mathbf{p}_T}^2]-J_1\omega_2 (M+m_1-{\omega_2})-J_2],\nonumber\\&&
A_{12}^{\mathcal{V},d}=A_{12}^{'\mathcal{V},d}=\frac{8}{3}[J_0 {\mathbf{p}_T}^2 {\mathbf{q}_T}^2+J_1 -({\mathbf{p}_T}^2+{\mathbf{q}_T}^2)+J_2],\nonumber\\&&
\end{eqnarray}

\begin{eqnarray}
&&A_{21}^{\mathcal{S}}= J_0,\nonumber\\
&&A_{21}^{\mathcal{S}}= J_0,\nonumber\\
&&A_{21}^{\mathcal{P}}=\frac{16}{3}J_0 (M+{\omega_1}) [{\mathbf{p}_T}^2\omega_1-m_1  {\mathbf{q}_T}^2]-\frac{32}{3}J_1 \omega_1 (M+{\omega_1})+\frac{16}{3{\mathbf{p}_T}^2}J_2
   (M+{\omega_1}) (m_1+{\omega_1}),\nonumber\\&&
A_{21}^{'\mathcal{P}}=\frac{16}{3}J_0\omega_2[{\mathbf{p}_T}^2 ({\omega_2}-M)-m_1 {\mathbf{q}_T}^2]-\frac{32}{3}J_1 {\omega_2 ({\omega_2}-M)+\frac{16}{3\mathbf{p}_T}^2}J_2
  \omega_2 (-M+m_1+{\omega_2}),
\nonumber\\&&
A_{21}^{\mathcal{V},a}=-J_0  [-4 {m_\mathcal{V}}^2\omega_1 (M+{\omega_1})+{m_\mathcal{V}}^2 {\mathbf{p}_T}^2+{m_\mathcal{V}}^2
   {\mathbf{q}_T}^2+({\mathbf{p}_T}^2-{\mathbf{q}_T}^2)^2]/{{m_\mathcal{V}}^2}-2J_1,
\nonumber\\&&
A_{21}^{'\mathcal{V},a}=-J_0  [4 {m_\mathcal{V}}^2\omega_2 (M-{\omega_2})+{m_\mathcal{V}}^2 {\mathbf{p}_T}^2+{m_\mathcal{V}}^2
   {\mathbf{q}_T}^2+({\mathbf{p}_T}^2-{\mathbf{q}_T}^2)^2]/{{m_\mathcal{V}}^2}-2J_1,
\nonumber\\&&
A_{21}^{\mathcal{V},b}=
A_{21}^{'\mathcal{V},b}=0,\nonumber\\&&
A_{21}^{\mathcal{V},c}=-4J_0  (M+{\omega_1}) (m_1+{\omega_1})+\frac{4}{{\mathbf{p}_T}^2}J_1 (M+{\omega_1}) (m_1+{\omega_1}),\nonumber\\&&
A_{21}^{'\mathcal{V},c}=-4J_0 \omega_2 (-M+m_1+{\omega_2})+\frac{4}{{\mathbf{p}_T}^2}J_1\omega_2 (-M+m_1+{\omega_2}),\nonumber\\&&
A_{21}^{\mathcal{V},d}=A_{21}^{'\mathcal{V},d}=\frac{-8}{3}J_0  \left({\mathbf{p}_T}^2+{\mathbf{q}_T}^2\right)+\frac{16}{3}J_1 ,\nonumber\\&&
\end{eqnarray}

\begin{eqnarray}
&&A_{22}^{\mathcal{S}}= (m_1+\omega_1)J_1/\mathbf{p}_T^2,\nonumber\\
&&A_{22}^{\mathcal{'S}}= (m_1+\omega_2-M)J_1/\mathbf{p}_T^2,\nonumber\\
&&A_{22}^{\mathcal{P}}=\frac{16}{3}\{J_0  {\mathbf{q}_T}^2 (M+{\omega_1}) [{\mathbf{p}_T}^2-2\omega_1 (m_1+{\omega_1})]J_1\omega_1 (M+{\omega_1})
   (m_1+{\omega_1})(1+{\mathbf{q}_T}^2/{\mathbf{p}_T}^2)\nonumber\\&&-J_2  (M+{\omega_1})\},\nonumber\\&&
A_{22}^{'\mathcal{P}}=\frac{16}{3}\{J_0\omega_2 {\mathbf{q}_T}^2[-2 M^2+2 M (m_1+2\omega_2)+ {\mathbf{p}_T}^2-2
  \omega_2 (m_1+{\omega_2})]\nonumber\\&&+J_1\omega_2  (1+{\mathbf{q}_T}^2/{\mathbf{p}_T}^2)[M^2 -M (m_1+2\omega_2)
   +{\omega_2} (m_1+{\omega_2})]-J_2 \omega_2\},
\nonumber\\&&
A_{22}^{\mathcal{V},a}=-J_1 (m_1+{\omega_1})[-4 {m_\mathcal{V}}^2\omega_1 (M+{\omega_1})+{m_\mathcal{V}}^2 {\mathbf{p}_T}^2+{m_\mathcal{V}}^2
   {\mathbf{q}_T}^2+\left({\mathbf{p}_T}^2-{\mathbf{q}_T}^2\right)^2]/({{m_\mathcal{V}}^2}{\mathbf{p}_T}^2)\nonumber\\&&-\frac{2}{{\mathbf{p}_T}^2}J_2 (m_1+{\omega_1}),
\nonumber\\&&
A_{22}^{'\mathcal{V},a}=J_1 (M-m_1-{\omega_2}) [4 {m_\mathcal{V}}^2\omega_2 (M-{\omega_2})+{m_\mathcal{V}}^2 {\mathbf{p}_T}^2+{m_\mathcal{V}}^2
   {\mathbf{q}_T}^2+\left({\mathbf{p}_T}^2-{\mathbf{q}_T}^2\right)^2]/({{m_\mathcal{V}}^2}{\mathbf{p}_T}^2)\nonumber\\&&+\frac{2}{{\mathbf{p}_T}^2}J_2 (M-m_1-{\omega_2}),
\nonumber\\&&
A_{22}^{\mathcal{V},b}=\frac{-4}{3}[J_0  {\mathbf{q}_T}^2 (m_1+{\omega_1})-J_2 (m_1+{\omega_1})/{\mathbf{p}_T}^2],
\nonumber\\&&
A_{22}^{'\mathcal{V},b}=\frac{-4}{3}[J_0 {\mathbf{q}_T}^2 (-M+m_1+{\omega_2})-J_2 (-M+m_1+{\omega_2})/{\mathbf{p}_T}^2 ],
\nonumber\\&&
A_{22}^{\mathcal{V},c}=-4[J_0 {\mathbf{q}_T}^2 (m_1 -M )+J_1 (M +\omega_1)+J_2
   (-m_1-{\omega_1})/{\mathbf{p}_T}^2],
\nonumber\\&&
A_{22}^{'\mathcal{V},c}=4[J_0 {\mathbf{q}_T}^2 (M-m_1)-J_1 \omega_2-J_2 (M-m_1-{\omega_2})/{\mathbf{p}_T}^2],
\nonumber\\&&
A_{22}^{\mathcal{V},d}=\frac{-8}{3}[J_0  {\mathbf{q}_T}^2 (m_1+{\omega_1})-J_1 (m_1+{\omega_1}) (1+{\mathbf{q}_T}^2/{\mathbf{p}_T}^2)+J_2 (m_1+{\omega_1})/{\mathbf{p}_T}^2],\nonumber\\&&
A_{22}^{'\mathcal{V},d}=\frac{-8}{3}[J_0  {\mathbf{q}_T}^2 (-M+m_1+{\omega_2})-J_1 (1+{\mathbf{q}_T}^2/{\mathbf{p}_T}^2) (-M+m_1+{\omega_2})\nonumber\\&&+J_2
   (-M+m_1+{\omega_2})/{\mathbf{p}_T}^2].\nonumber\\&&
\end{eqnarray}

\end{document}